\documentclass[prd,preprint,preprintnumbers,nofootinbib,eqsecnum,superscriptaddress]{revtex4}

 \usepackage[dvips,final]{graphicx}
  \usepackage{amssymb}
   \usepackage{amsmath}
    \usepackage{amsfonts}
     \usepackage{epsfig}
      \usepackage{bm}

\usepackage{enumitem}
\usepackage[normalem]{ulem}
\usepackage{mathpazo}

\usepackage{multirow}
\usepackage{ctable}
\usepackage{booktabs}
\usepackage{array}
\usepackage{tabularx}
\usepackage{xcolor}
\usepackage{pstricks}

\usepackage[section]{placeins}

\input epsf.tex
\def\desepsf(#1 width #2){\epsfxsize=#2 \epsfbox{#1}}

\usepackage[normalem]{ulem}

\usepackage{multirow}
\usepackage{ctable}
\usepackage{booktabs}
\usepackage{array}
\usepackage{tabularx}
\usepackage{xcolor}
\usepackage{pstricks}
\definecolor{fred}{rgb}{0.90053, 0.00369, 0.00159}  

\newcommand{\be}{\begin{eqnarray}}
\newcommand{\ee}{\end{eqnarray}}

\begin{document}

\author{Rafa{\l} Maciu{\l}a}
\email{rafal.maciula@ifj.edu.pl}
\affiliation{Institute of Nuclear
Physics, Polish Academy of Sciences, ul. Radzikowskiego 152, PL-31-342 Krak{\'o}w, Poland}

\author{Antoni Szczurek}
\email{antoni.szczurek@ifj.edu.pl} 
\affiliation{Institute of Nuclear Physics, Polish Academy of Sciences, ul. Radzikowskiego 152, PL-31-342 Krak{\'o}w, Poland}
\affiliation{College of Natural Sciences, Institute of Physics, University of Rzesz{\'o}w, ul. Pigonia 1, PL-35-310 Rzesz{\'o}w, Poland}

\title{Far-forward production of charm mesons and neutrinos
at Forward Physics Facilities at the LHC
 and the intrinsic charm in the proton}

\begin{abstract}
We discuss production of far-forward charm/anticharm quarks,
$D$ mesons/antimesons and neutrinos/antineutrinos from their semileptonic decays
in proton-proton collisions at the LHC energies. We include the gluon-gluon fusion $gg \to c\bar{c}$, the intrinsic charm (IC) $gc \to gc$ as well as the recombination $gq \to Dc$ partonic mechanisms. The calculations are performed within the $k_T$-factorization approach and the hybrid model using different unintegrated parton distribution functions (uPDFs) for gluons from the literature, as well as within the collinear approach. We compare our results to the LHCb data for forward $D^{0}$-meson production at $\sqrt{s} = 13$ TeV for different
rapidity bins in the interval $2 < y <  4.5$. A good description is achieved for the Martin-Ryskin-Watt (MRW) uPDF. 
We also show results for the Kutak-Sapeta (KS) gluon uPDF, both in the linear form and including nonlinear effects. The nonlinear effects play a role only at very small transverse momenta of $D^0$ or $\bar D^0$ mesons. The IC and recombination models are negligible at the LHCb kinematics. Both the mechanisms start to be crucial at larger rapidities and dominate over the standard charm production mechanisms.
At high energies there are so far no experiments probing this region. We present uncertainty bands for the both mechanisms. 
Decreased uncertainty bands will be available soon from fixed-target charm 
experiments in $pA$-collisions.
We present also energy distributions for forward electron, muon and tau 
neutrinos to be measured at the LHC by the currently operating FASER$\nu$ experiment, as well as by future experiments like FASER$\nu2$ or FLArE, proposed very recently by the Forward Physics Facility project.

Again components of different mechanisms are shown separately.
For all kinds of neutrinos (electron, muon, tau) the subleading
contributions, i.e. the IC and/or the recombination, dominate over light meson
(pion, kaon) and the standard charm production contribution driven by fusion of gluons for neutrino energies $E_{\nu} \gtrsim 300$ GeV.
For electron and muon neutrinos both the mechanisms lead to a similar production rates and their separation seems rather impossible. 
On the other hand, for $\nu_{\tau} + {\bar \nu}_{\tau}$ neutrino flux
the recombination is further reduced making the measurement
of the IC contribution very attractive.
\end{abstract} 

\maketitle

\section{Introduction}

At high energies a process of mid-rapidity production of charm quark/antiquarks is dominated 
by fusion of gluons (pair creation mechanisms) and interactions of gluons with light quarks (flavour excitation mechanisms).
This process can be well described at a broad energy range within theoretical models based on either the next-to-leading order (NLO) collinear or the $k_{T}$-factorization frameworks (see e.g. Refs.\cite{Maciula:2013wg,Maciula:2015kea,Cacciari:2012ny,Kniehl:2012ti,Klasen:2014dba}). The forward production of charm is not fully under control. There are some mechanisms which may play a role outside the mid-rapidity region (forward/backward production) not only at high collision energies.
There are potentially two QCD mechanisms that may play a role in this
region:
\begin{enumerate}[label=(\alph*)]
\item the mechanism of production of charm initiated by intrinsic charm
which can be called knock-out of the intrinsic charm and
\item recombination of charm quarks/antiquarks and light
antiquarks/quarks.
\end{enumerate}

Recently we have shown, that at lower energies the mechanisms easily mix and it is difficult to
disentangle them in the backward production of $D$ mesons \cite{Maciula:2022otw}.
Nevertheless such fixed-target experiments provide some limitations
on the not fully explored mechanisms. The associated uncertainties are not small. 
The asymmetry in the production of $D^0$ ($\bar D^0$) mesons may soon
provide an interesting information on the recombination mechanism.
Some limitations on the intrinsic charm component were obtained
recently based on the IceCube neutrino data \cite{Goncalves:2021yvw} (see also Ref.~\cite{Laha:2016dri}).
From such experiments we get roughly the probability to find
intrinsic charm in the nucleon $P_{IC} <$ 1 \%, which is consistent with the central value of the CT14nnloIC PDF global fit \cite{Hou:2017khm}
as well as with the recent study of the NNPDF group based on machine
learning and a large experimental dataset \cite{Ball:2022qks}.

In this paper we will discuss far-forward production
of charm at the LHC energies. The forward production of
charm quarks is associated with forward production of charmed mesons.
Their direct measurement in far-forward directions is challenging.
The forward production of mesons leads to forward production
of neutrinos coming from their semileptonic decays. So far such neutrinos cannot be measured at the LHC, however,
there are some ongoing projects to improve the situation.
Recently several new detectors were proposed to measure
the forward neutrinos (e.g. FASER$\nu$, SND@LHC, FASER$\nu2$, FLArE), according to the Forward Physics Facility (FPF) proposal \cite{Anchordoqui:2021ghd,FASER:2019dxq,SNDLHC:2022ihg}.
Here we wish to summarize the situation for the collider mode of the LHC.
Can such measurements provide new interesting information on the poorly
known mechanisms? We will try to answer the question.

In principle, our present study extends predictions for the far-forward neutrino fluxes at the LHC reported recently in Refs.~\cite{Bai:2020ukz,Kling:2021gos,Bai:2021ira} by taking into account new mechanisms that have not been considered so far in this context.

\section{Details of the model calculations}

In the present study we take into consideration three different
production mechanisms of charm, including:
\begin{enumerate}[label=(\alph*)]
\item the standard (and usually considered as a leading) QCD mechanism of gluon-gluon fusion: $g^*g^* \to
c\bar c$ with off-shell initial state partons, calculated both in the full $k_{T}$-factorization approach and in the hybrid model;
\item the mechanism driven by the intrinsic charm component of
proton: $g^*c \to gc$ calculated in the hybrid approach with off-shell initial state gluon and collinear intrinsic charm distribution;
\item the recombination mechanism: $gq \to Dc$, calculated in the leading-order collinear approach.
\end{enumerate}
Calculations of the three contributions are performed following
our previous studies reported in Refs.~\cite{Maciula:2019izq,Maciula:2020cfy,Maciula:2020dxv,Maciula:2021orz,Maciula:2022otw}.

\subsection{The standard QCD mechanism for charm production}

\begin{figure}[!h]
\centering
\begin{minipage}{0.3\textwidth}
  \centerline{\includegraphics[width=1.0\textwidth]{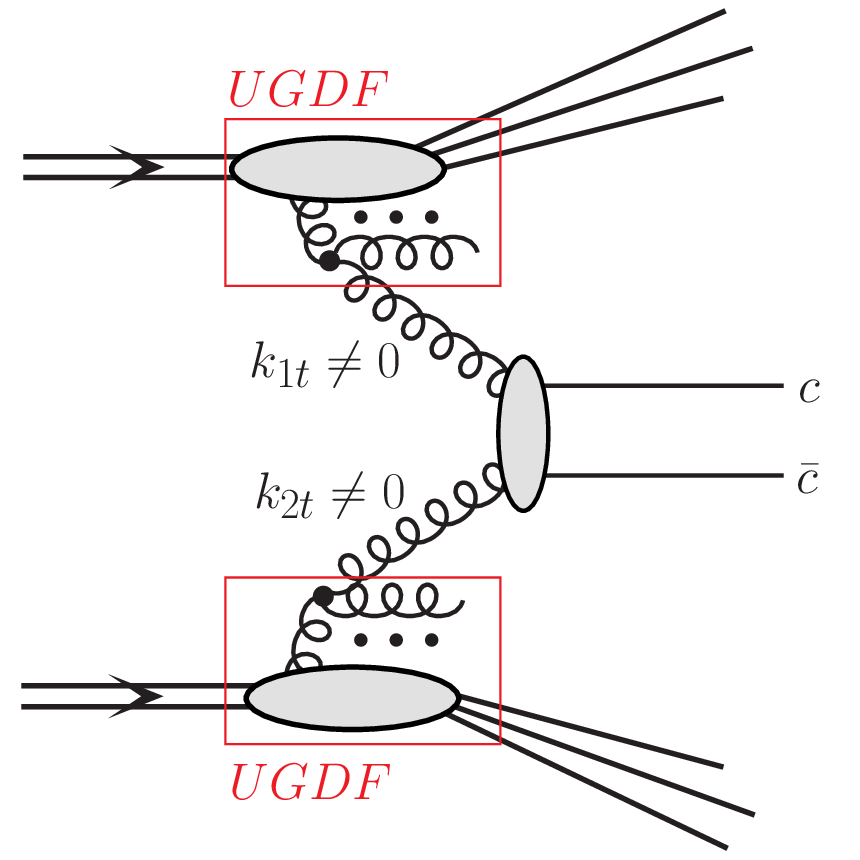}}
\end{minipage}
  \caption{
\small A diagram of the standard QCD mechanism of charm production in the $k_{T}$-factorization approach driven by the fusion of two off-shell gluons.
}
\label{fig:diagramLO}
\end{figure}

Here we follow the theoretical formalism for
the calculation of the $c\bar{c}$-pair production in the
$k_{T}$-factorization approach \cite{kTfactorization}. In this framework the transverse momenta
$k_{t}$'s (or virtualities) of both partons entering the hard process are taken into account, both in the matrix elements and in the parton distribution functions. Emission of the initial state partons is encoded in the transverse-momentum-dependent (unintegrated) PDFs (uPDFs). In the case of charm flavour production the parton-level cross section is usually calculated via the $2\to 2$ leading-order $g^*g^* \to c\bar c$ fusion mechanism with off-shell initial state gluons that is the dominant process at high energies (see Fig.~\ref{fig:diagramLO}), especially in the mid-rapidity region. However, even at extremely backward/forward rapidities the $q^*\bar q^* \to c\bar c $ mechanism remains subleading. Then the hadron-level differential cross section for the $c \bar c$-pair production, formally at leading-order, reads:
\begin{eqnarray}\label{LO_kt-factorization} 
\frac{d \sigma(p p \to c \bar c \, X)}{d y_1 d y_2 d^2p_{1,t} d^2p_{2,t}} &=&
\int \frac{d^2 k_{1,t}}{\pi} \frac{d^2 k_{2,t}}{\pi}
\frac{1}{16 \pi^2 (x_1 x_2 s)^2} \; \overline{ | {\cal M}^{\mathrm{off-shell}}_{g^* g^* \to c \bar c} |^2}
 \\  
&& \times  \; \delta^{2} \left( \vec{k}_{1,t} + \vec{k}_{2,t} 
                 - \vec{p}_{1,t} - \vec{p}_{2,t} \right) \;
{\cal F}_g(x_1,k_{1,t}^2,\mu_{F}^2) \; {\cal F}_g(x_2,k_{2,t}^2,\mu_{F}^2) \; \nonumber ,   
\end{eqnarray}
where ${\cal F}_g(x_1,k_{1,t}^2,\mu_{F}^2)$ and ${\cal F}_g(x_2,k_{2,t}^2,\mu_{F}^2)$
are the gluon uPDFs for both colliding hadrons and ${\cal M}^{\mathrm{off-shell}}_{g^* g^* \to c \bar c}$ is the off-shell matrix element for the hard subprocess.
The gluon uPDF depends on gluon longitudinal momentum fraction $x$, transverse momentum
squared $k_t^2$ of the gluons entering the hard process, and in general also on a (factorization) scale of the hard process $\mu_{F}^2$. They must be evaluated at longitudinal momentum fractions 
$x_1 = \frac{m_{1,t}}{\sqrt{s}}\exp( y_1) + \frac{m_{2,t}}{\sqrt{s}}\exp( y_2)$, and $x_2 = \frac{m_{1,t}}{\sqrt{s}}\exp(-y_1) + \frac{m_{2,t}}{\sqrt{s}}\exp(-y_2)$, where $m_{i,t} = \sqrt{p_{i,t}^2 + m_c^2}$ is the quark/antiquark transverse mass. 

As we have carefully discussed in Ref.~\cite{Maciula:2019izq}, there is a direct relation between a resummation present in uPDFs in the transverse momentum dependent factorization and a parton shower in the collinear framework. In most uPDF the off-shell gluon can be produced either from gluon or quark, therefore, in the $k_{T}$-factorization all channels of the higher-order type in the collinear approach driven by $gg, q\bar q$ and even by $qg$ initial states are open already at leading-order (in contrast to the collinear factorization).

In the numerical calculations below we apply the Martin-Ryskin-Watt (MRW) \cite{Watt:2003mx} gluon uPDFs calculated from the MMHT2014nlo \cite{Harland-Lang:2014zoa} collinear gluon PDF as well as Kutak-Sapeta (KS) \cite{Kutak:2014wga} linear and nonlinear distributions. As a default set in the numerical calculations we take the renormalization scale
$\mu^2 = \mu_{R}^{2} = \sum_{i=1}^{n} \frac{m^{2}_{it}}{n}$ (averaged
transverse mass of the given final state) and the charm quark mass
$m_{c}=1.5$ GeV. The strong-coupling constant $\alpha_{s}(\mu_{R}^{2})$
at next-to-next-to-leading-order is taken from the CT14nnloIC PDF \cite{Hou:2017khm}
routines.

In the parts of the calculations especially devoted to the far-forward production of charm we also calculate the standard production mechanism with only one gluon being off-shell and second one collinear in accordance with the assumptions of the so called hybrid model, which is described in the next subsection.

\subsection{The Intrinsic Charm induced component}

\begin{figure}[!h]
\centering
\begin{minipage}{0.3\textwidth}
  \centerline{\includegraphics[width=1.0\textwidth]{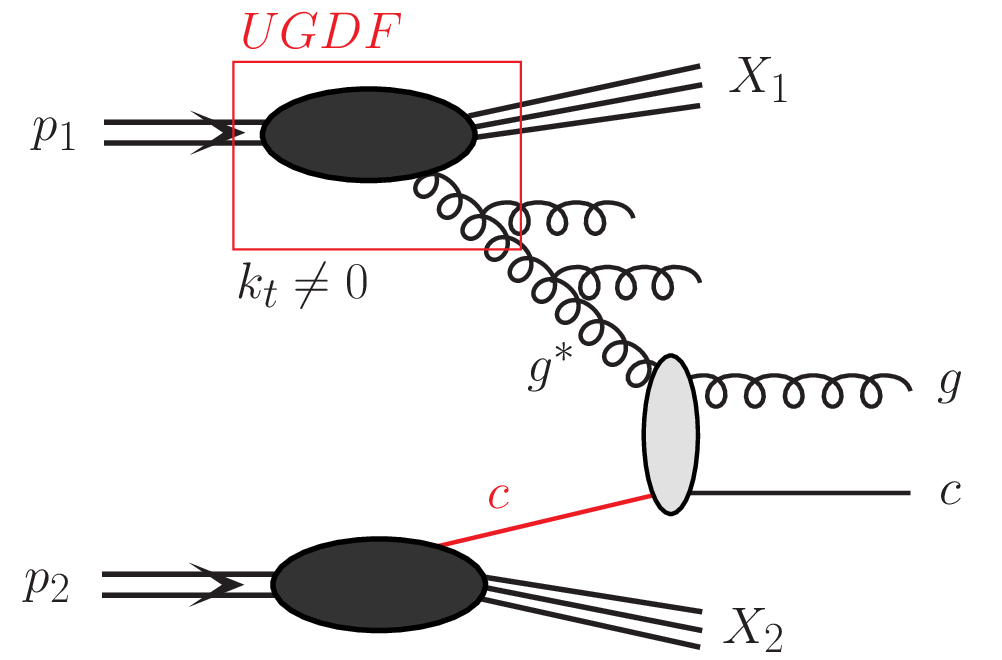}}
\end{minipage}
  \caption{
\small A diagrammatic representation of the intrinsic charm driven mechanism of charm production within the hybrid model with the off-shell gluon and the on-shell charm quark in the initial state.
}
\label{fig:diagramIC}
\end{figure}

The intrinsic charm contribution to charm production cross section (see
Fig.~\ref{fig:diagramIC}) is obtained within the hybrid theoretical
model discussed by us in detail in Ref.~\cite{Maciula:2020dxv}. The FPF experiments at the LHC
will allow to explore the charm cross section in
the far-forward rapidity direction where an asymmetric kinematical
configurations are selected. Thus in the basic $gc \to gc$ reaction the gluon PDF and the intrinsic charm PDF are simultaneously probed at different longitudinal momentum fractions - extremely small for the gluon and very large for the charm quark.

Within the asymmetric kinematic configuration $x_1 \ll x_2$ the cross section for the processes under consideration can be calculated in the so-called hybrid factorization model motivated by the work in Ref.~\cite{Deak:2009xt}. In this framework the small-$x$ gluon is taken to be off mass shell and 
the differential cross section e.g. for $pp \to g c X$ via $g^* c \to g c$ mechanism reads:
\begin{eqnarray}
d \sigma_{pp \to gc X} = \int d^ 2 k_{t} \int \frac{dx_1}{x_1} \int dx_2 \;
{\cal F}_{g^{*}}(x_1, k_{t}^{2}, \mu^2) \; c(x_2, \mu^2) \; d\hat{\sigma}_{g^{*}c \to gc} \; ,
\end{eqnarray}
where ${\cal F}_{g^{*}}(x_1, k_{t}^{2}, \mu^2)$ is the unintegrated
gluon distribution in one proton and $c(x_2, \mu^2)$ a collinear PDF in
the second one. The $d\hat{\sigma}_{g^{*}c \to gc}$ is the hard partonic
cross section obtained from a gauge invariant tree-level off-shell
amplitude. A derivation of the hybrid factorization from the dilute
limit of the Color Glass Condensate approach can be found e.g. in Ref.~\cite{Dumitru:2005gt} (see also Ref.~\cite{Kotko:2015ura}). The relevant cross sections are calculated with the help of the \textsc{KaTie} Monte Carlo generator \cite{vanHameren:2016kkz}. There the initial state quarks (including heavy quarks) can be treated as a massless partons only.  

Working with minijets (jets with transverse momentum of the order of a few GeV) requires a phenomenologically motivated regularization of the cross sections. Here we follow the minijet model \cite{Sjostrand:1987su} adopted e.g. in \textsc{Pythia} Monte Carlo generator, where a special suppression factor is introduced at the cross section level \cite{Sjostrand:2014zea}:
\begin{equation}
F(p_t) = \frac{p_t^2}{ p_{T0}^2 + p_t^2 } \; 
\label{Phytia_formfactor}
\end{equation}
for each of the outgoing massless partons with transverse momentum $p_t$, where $p_{T0}$ is a free parameter of the form factor
that also enters as an argument of the strong coupling constant $\alpha_{S}(p_{T0}^2+\mu_{R}^{2})$. A phenomenological motivation behind its application in the $k_{T}$-factorization approach is discussed in detail in Ref.~\cite{Kotko:2016lej}.



In the numerical calculations below, the intrinsic charm PDFs are taken at the initial scale $m_{c} = 1.3$ GeV, so the perturbative charm contribution is intentionally not taken into account. We apply different grids of the intrinsic charm distribution from the CT14nnloIC PDF \cite{Hou:2017khm} that correspond to the BHPS model \cite{BHPS1980}.    

\subsection{Recombination model of charmed meson production}

\begin{figure}[!h]
\includegraphics[width=5cm]{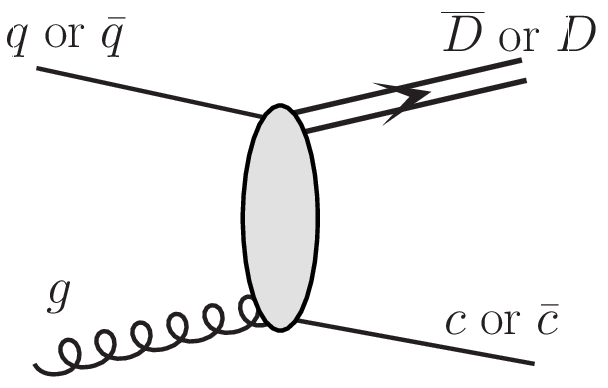}
\hskip+5mm
\includegraphics[width=5cm]{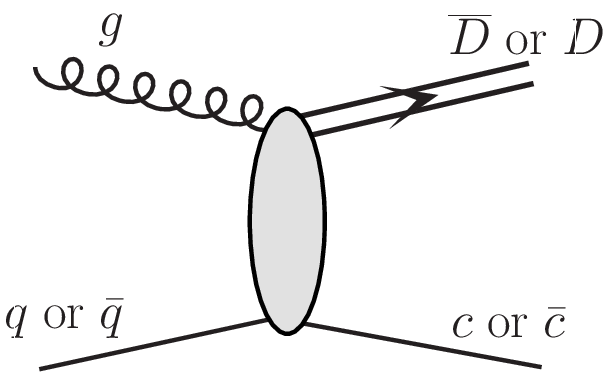}
\caption{Generic leading-order diagrams for $D$ meson production via 
the BJM recombination.}
\label{fig:recombination_diagrams}
\end{figure}

The underlying mechanism of the  Braaten-Jia-Mechen (BJM) \cite{BJM2002a,BJM2002b,BJM2002c} recombination is illustrated 
in Fig.~\ref{fig:recombination_diagrams}. Differential cross section for production of $D c$ final state reads:
\begin{eqnarray}
\frac{d\sigma}{d y_1 d y_2 d^2 p_{t}} = \frac{1}{16 \pi^2 {\hat s}^2}
&& [ x_1 q_1(x_1,\mu^2) \, x_2 g_2(x_2,\mu^2)
\overline{ | {\cal M}_{q g \to D c}(s,t,u)|^2} \nonumber \\
&+& x_1 g_1(x_1,\mu^2) \, x_2 q_2(x_2,\mu^2)
\overline{ | {\cal M}_{g q \to D c}(s,t,u)|^2} ]  \, .
\label{cross_section}
\end{eqnarray}
Above $y_1$ is rapidity of the $D$ meson and $y_2$ rapidity of 
the associated $c$ or $\bar c$. The fragmentation of the latter
will be discussed below.

The matrix element squared in (\ref{cross_section}) can be written as
\begin{equation}
\overline{ | {\cal M}_{q g \to D c}(s,t,u)|^2} =
\overline{ | {\cal M}_{q g \to ({\bar c} q)^n c} |^2} \cdot \rho \; ,
\label{rho_definition}
\end{equation}
where $n$ enumerates quantum numbers of the ${\bar c} q$ system
$n \equiv ^{2J+1}L$ and
$\rho$ can be interpreted as a probability to form real meson.
For illustration as our default set we shall take $\rho$ = 0.1, but the precise number
should be adjusted to experimental data. For the discussion of
the parameter see e.g. Refs.~\cite{BJM2002b,BJM2002c} and references therein.
The asymmetries observed in photoproduction can be explained with $\rho$ = 0.15
\cite{BJM2002c}. Some constrains for this parameter could be also obtained from
the LHCb fixed-target data on $D$-meson production asymmetry that are going to be published soon \cite{Maciula:2022otw}. 

The explicit form of the matrix element squared can be found in
\cite{BJM2002a} for pseudoscalar and vector meson production for color
singlet and color octet meson-like states. Similar formula can be written for production of ${\bar D} {\bar c}$.
Then the quark distribution is replaced by the antiquark distribution.
In the following we include only color singlet $(q {\bar c})^n$
or $({\bar q} c)^n$ components.
As a default set, the factorization scale in the calculation is taken as:
\begin{equation}
\mu^2 = p_t^2 + \frac{m_{t,D}^2 + m_{t,c}^2}{2} \; .
\label{scale}
\end{equation}

Within the recombination mechanism we include fragmentation of $c$-quarks or $\bar{c}$-antiquarks accompanying directly produced
$D$-mesons or $\overline{D}$-antimesons, e.g.:
\begin{eqnarray}
d \sigma [q g \to \overline{D}_{\mathrm{direct}} + D_{\mathrm{frag.}}] &=& d \sigma [q g \to \overline{D} + c] 
\otimes F^{\mathrm{frag.}}_{c \to D} \; ,
\label{fragmentation}
\end{eqnarray}
where $F^{\mathrm{frag.}}_{c \to D}$ is the relevant fragmentation function.
How the convolution $\otimes$ is understood
is explained in \cite{Maciula:2019iak}.

We shall discuss in the present paper also the asymmetry in production of
$D^0$ meson and ${\overline D}^0$ antimeson.
The asymmetry is defined as:
\begin{equation}
A_{p} = \frac{d\sigma^{D^0}\!\!/d\xi - d\sigma^{\overline{D}^0}\!\!/d\xi}
                   {d\sigma^{D^0}\!\!/d\xi + d\sigma^{\overline{D}^0}\!\!/d\xi}
\; ,
\label{asymmetry}
\end{equation}
where $\xi$ represents single variable ($y$ or $p_t$) or even a pair of
variables (${y, p_t}$).

Only a part of the pseudoscalar $D$ mesons is directly produced.
A second part originates from vector meson decays.
The vector $D$ mesons promptly and dominantly decay to pseudoscalar mesons:
\begin{eqnarray}
D^{*0} &\to& D^0 \pi (0.619), D^0 \gamma (0.381) , \nonumber \\ 
D^{*+} &\to& D^0 \pi^+ (0.677), D^+ \pi^0 (0.307), D^+ \gamma (0.0016) , \nonumber \\
D_s^{*+} &\to& D_s^+ \gamma (0.935), D_s^+ \pi^0 (0.058), D_s^+ e^+ e^- (6.7 10^{-3}).
\end{eqnarray}
%

%
%

\subsection{Hadronization of charm quarks}

The transition of charm quarks to open charm mesons is done in the
framework of the independent parton fragmentation picture (see
\textit{e.g.} Refs.~\cite{Maciula:2019iak}) where the inclusive
distributions of open charm meson can be obtained through a convolution
of inclusive distributions of produced charm quarks/antiquarks and $c \to D$ fragmentation functions. Here we follow exactly the method which was applied by us in our previous study of forward/backward charm production reported e.g. in Ref.~\cite{Maciula:2021orz}. According to this approach we assume that the $D$-meson is emitted in the direction of parent $c$-quark/antiquark, i.e. $\eta_D=\eta_c$ (the same pseudorapidities or polar angles) and the $z$-scaling variable is defined with the light-cone momentum i.e. $p^{+}_{c} = \frac{p^{+}_{D}}{z}$ where $p^{+} = E + p$. In numerical calculations we take the Peterson fragmentation function \cite{Peterson:1982ak} with $\varepsilon = 0.05$, often used in the context of hadronization of heavy flavours. Then, the hadronic cross section is normalized by the relevant charm fragmentation fractions for a given type of $D$ meson \cite{Lisovyi:2015uqa}. In the numerical calculations below when discussing $D^0$-meson production for $c \to D^{0}$ transition we take the fragmentation probability $\mathrm{P}_{c \to D^{0}} = 61\%$.

\subsection{Production of $\nu_{e}$ and $\nu_{\mu}$ neutrinos}

There are different sources of neutrinos (see \cite{ParticleDataGroup:2022pth}). In general, the $\nu_{e}$ neutrinos can be produced from the decays of $K^{+}$ and $K_{L}$ mesons and the $\nu_{\mu}$ neutrinos from $K^{+}$, $K_{L}$ and $\pi^{+}$. In addition both of them can be also produced from $D^{+}, D^{0}, D_{S}^{+}$ mesons via many decay channels. Another important source of $\nu_{e}$ and $\nu_{\mu}$ is also decay of muons. The latter component is important for large distances, $c \tau$ = 659 m \cite{ParticleDataGroup:2022pth}.
The planned neutrino target (the target where neutrinos are measured) will be placed 480 m from the interaction point. 
This requires more dedicated studies. In addition, there are many sources of muons.
Realiable estimation would require evolution of produced neutrinos through the rock between the production point and analysing target.
All the most important decay channels for $\nu_{e}$ and $\nu_{\mu}$ neutrinos are collected in Table \ref{tab:decays}. In the present study we are particularly interested in $D$ meson leptonic and semileptonic decays.

\begin{table}[tb]%
\caption{Leading decay channels resulting in production of $\nu_{\mu}$ and $\nu_{e}$ neutrinos.}

\label{tab:decays}
\centering %
\begin{tabularx}{0.85\linewidth}{c c c}
\\[-4.ex] 
\toprule[0.1em] %
\\[-4.ex] 

\multirow{1}{4.5cm}{Neutrino source} & \multirow{1}{6.cm}{Leading decay channels} & \multirow{1}{2.cm}{BR [\%]}\\ [+0.1ex]
\bottomrule[0.1em]
\multirow{1}{4.5cm}{charged pions} & \multirow{1}{6.cm}{$\pi^+ \to \mu^+ \nu_{\mu}$} & \multirow{1}{2.cm}{99.99}\\ [+0.1ex]
\hline
\multirow{1}{4.5cm}{kaons: $K^{+}, K_{L}$} & \multirow{1}{6.cm}{ $K^{+} \to \mu^+ \nu_{\mu}$} & \multirow{1}{2.cm}{63.56}\\ [-1.5ex]
\multirow{1}{4.5cm}{} & \multirow{1}{6.cm}{$K^{+} \to \pi^0 \mu^+ \nu_{\mu}$} & \multirow{1}{2.cm}{3.35}\\ [-1.5ex]
\multirow{1}{4.5cm}{} & \multirow{1}{6.cm}{$K^{+} \to \pi^0 e^+ \nu_e$} & \multirow{1}{2.cm}{5.07}\\ [-1.5ex]
\multirow{1}{4.5cm}{} & \multirow{1}{6.cm}{$K^{+} \to e^+ \nu_{e}$} & \multirow{1}{2.cm}{$1.58 \cdot 10^{-5}$}\\ [-1.5ex]
\multirow{1}{4.5cm}{} & \multirow{1}{6.cm}{$K_L \to \pi^{\pm} \mu^{\mp} \nu_{\mu}$ } & \multirow{1}{2.cm}{$27.04$}\\ [-1.5ex]
\multirow{1}{4.5cm}{} & \multirow{1}{6.cm}{$K_L \to \pi^{\pm} e^{\mp} \nu_e$} & \multirow{1}{2.cm}{$40.55$}\\ [+0.1ex]
\hline
\multirow{1}{4.5cm}{charm mesons:} & \multirow{1}{6.cm}{$D^{+} \to {\bar K}^0 \mu^+ \nu_{\mu}$} & \multirow{1}{2.cm}{8.76}\\ [-1.5ex]
\multirow{1}{4.5cm}{$D^{+}, D^{0}, D^{+}_{S}$} & \multirow{1}{6.cm}{$D^{+} \to K^- \pi^+ \mu^+ \nu_{\mu}$} & \multirow{1}{2.cm}{3.65}\\ [-1.5ex]
\multirow{1}{4.5cm}{} & \multirow{1}{6.cm}{$D^{+} \to {\bar K}^0 e^+ \nu_e$} & \multirow{1}{2.cm}{$8.72$}\\ [-1.5ex]
\multirow{1}{4.5cm}{} & \multirow{1}{6.cm}{$D^{+} \to K^- \pi^+ e^+ \nu_e$ } & \multirow{1}{2.cm}{$4.02$}\\ [-1.5ex]
\multirow{1}{4.5cm}{} & \multirow{1}{6.cm}{$D^0 \to K^- \mu^+ \nu_{\mu}$} & \multirow{1}{2.cm}{3.41}\\ [-1.5ex]
\multirow{1}{4.5cm}{} & \multirow{1}{6.cm}{$D^0 \to K^*(892) \mu^+ \nu_{\mu}$} & \multirow{1}{2.cm}{1.89}\\ [-1.5ex]
\multirow{1}{4.5cm}{} & \multirow{1}{6.cm}{$D^0 \to K^- e^+ \nu_e$} & \multirow{1}{2.cm}{3.55}\\ [-1.5ex]
\multirow{1}{4.5cm}{} & \multirow{1}{6.cm}{$D^0 \to {\bar K}^0 \pi^- e^+ \nu_e$} & \multirow{1}{2.cm}{1.44}\\ [-1.5ex]
\multirow{1}{4.5cm}{} & \multirow{1}{6.cm}{$D^0 \to K^*(892) e^+ \nu_e$} & \multirow{1}{2.cm}{2.15}\\ [-1.5ex]
\multirow{1}{4.5cm}{} & \multirow{1}{6.cm}{$D^0 \to K^- \pi^0 e^+ \nu_e$} & \multirow{1}{2.cm}{1.60}\\ [-1.5ex]
\multirow{1}{4.5cm}{} & \multirow{1}{6.cm}{$D_s^{+} \to \eta e^+ \nu_e$ } & \multirow{1}{2.cm}{2.32}\\ [-1.5ex]
\multirow{1}{4.5cm}{} & \multirow{1}{6.cm}{$D_s^{+} \to \eta \mu^+ \nu_{\mu}$ } & \multirow{1}{2.cm}{2.40}\\ [-1.5ex]
\multirow{1}{4.5cm}{} & \multirow{1}{6.cm}{$D_s^{+} \to \phi e^+ \nu_e$ } & \multirow{1}{2.cm}{2.39}\\ [-1.5ex]
\multirow{1}{4.5cm}{} & \multirow{1}{6.cm}{$D_s^{+} \to \phi \mu^+ \nu_{\mu}$ } & \multirow{1}{2.cm}{1.9}\\ [-1.5ex]
\multirow{1}{4.5cm}{} & \multirow{1}{6.cm}{$D_s^{+} \to \eta' e^+ \nu_e$ } & \multirow{1}{2.cm}{$8.00 \cdot 10^{-3} $}\\ [-1.5ex]
\multirow{1}{4.5cm}{} & \multirow{1}{6.cm}{$D_s^{+} \to \eta' \mu^+ \nu_{\mu}$ } & \multirow{1}{2.cm}{1.01}\\ [+0.1ex]
\hline
\multirow{1}{4.5cm}{muons} & \multirow{1}{6.cm}{$\mu^{-} \to e^{-} \overline{\nu}_e \nu_{\mu}$} & \multirow{1}{2.cm}{$\approx$ 100}\\ [+0.1ex]
\bottomrule[0.1em]
\end{tabularx}
\end{table}

In the present study we are particularly interested in $D$ meson semileptonic decays. As will be discussed below we have no such decay functions. In practical evaluation often a simplified decay function for kaon decays \cite{Lipari:1993hd} is used also to the decays of charm mesons $D \to M_{\mathrm{eff}}\; l^+ \nu_l$, where $M_{\mathrm{eff}}$ is the effective invariant mass square in the decay,  
by replacing $m_K \to m_D$ in the simplified formula:
\begin{equation}
\label{lipari-formula}
G(x) = \frac{12 x^2 (1-\epsilon^2-x)^2}{g(\epsilon)(1-x)} \; ,
\end{equation}
where $g(\epsilon) = 1 - 8\epsilon^2 - 24\epsilon^4\ln\epsilon + 8\epsilon^6 - \epsilon^8$ with $\epsilon = M_{\mathrm{eff}}/m_D$, and $x =2 E_{\nu}^*/m_D$ with kinematical limits $0 \leq x \leq 1-\epsilon^2$.

By fitting the data in Ref.~\cite{Bugaev:1998bi} the authors find:
$M_{\mathrm{eff}}$ = 0.63 GeV for $D^{\pm}$, and
$M_{\mathrm{eff}}$ = 0.67 GeV for $D^0/{\bar D}^0$.
Almost the same number for both species of D mesons.
Such a form is not completely correct as there are several final
channels with neutrino/antineutrino as discussed above.

In future one could use also more theoretically motivated $D$-meson 
decay functions obtained from semileptonic transition form factors
(see e.g. Ref.~\cite{Yao:2020vef})
or use directly experimental data for semileptonic transition form factor
measured last years by the BESIII collaboration \cite{Yang:2018qdx}.

An alternative way to incorporate semileptonic decays into theoretical
model is to take relevant experimental input. Here we follow the method 
described in Refs.~\cite{Luszczak:2008je,Maciula:2015kea,Bolzoni:2012kx}. For example, the CLEO 
\cite{CLEO:2006ivk} collaboration has measured very precisely 
the momentum spectrum of electrons/positrons coming from the decays 
of $D$ mesons. This is done by producing resonances: $\Psi(3770)$
which decays into $D$ and $\bar D$ mesons.

This less ambitious but more pragmatic approach is based on purely 
empirical fits to (not absolutely normalized) CLEO experimental data 
points. These electron decay functions should account for the proper
branching fractions which are known experimentally 
(see e.g. \cite{ParticleDataGroup:2022pth,CLEO:2006ivk}).
The branching fractions for various species of $D$ mesons are different:
\begin{eqnarray}
&&\mathrm{BR}(D^+\to~e^+ \nu_e X)=16.13\pm 0.20(\mathrm{stat.})\pm 0.33(\mathrm{syst.})\%,\nonumber \\
&&\mathrm{BR}(D^0\to~e^+ \nu_e X)=6.46\pm 0.17(\mathrm{stat.})\pm 0.13(\mathrm{syst.})\%. 
\end{eqnarray}
Because the shapes of positron spectra for both decays are 
identical within error bars we can take the average value (for $D^{\pm}$
and $D^0$) of 
BR($D\!\to\!e \, \nu_e \, X) \approx 10 \%$ and 
simplify the calculation.

After renormalizing to experimental branching fractions the adjusted decay function is then used to generate leptons 
in the rest frame of the decaying $D$ meson in a Monte Carlo approach. In the case of semileptonic decays of $D$ mesons, relevant for
electron and muon neutrinos/antineutrinos, we generate 1000
decays for each considered (generated) $D$ meson. This way one can avoid all uncertainties associated with explicit calculations of semileptonic decays of mesons.

The open charm mesons are almost at rest, so in practice one measures the meson rest frame
distributions of electrons/positrons. With this assumption one can find a good fit to the CLEO data with:
\begin{eqnarray}
f^{Lab}_{CLEO}(p) &=& 12.55 (p+0.02)^{2.55} (0.98-p)^{2.75}\; . 
\label{CLEO_fit_function}
\end{eqnarray}
In these purely empirical parametrizations $p$ must be taken in GeV.

In order to take into account the small effect of the non-zero motion of the $D$ mesons in 
the case of the CLEO experiment, the above parametrization of the fit in the laboratory frame has to be modified. The improvement can be achieved by including the boost of the new modified rest frame functions to the CLEO laboratory frame. The quality of fit from Eqs.~(\ref{CLEO_fit_function}) will be reproduced. The $D$ rest frame decay function takes the following form:
\begin{eqnarray}
f^{Rest}_{CLEO}(p) &=& 12.7 (p+0.047)^{2.72} (0.9-p)^{2.21}\; . 
\label{CLEO_fit_boost}  \\
\end{eqnarray}
%

\begin{figure}[tb]
\begin{minipage}{0.47\textwidth}
 \centerline{\includegraphics[width=1.0\textwidth]{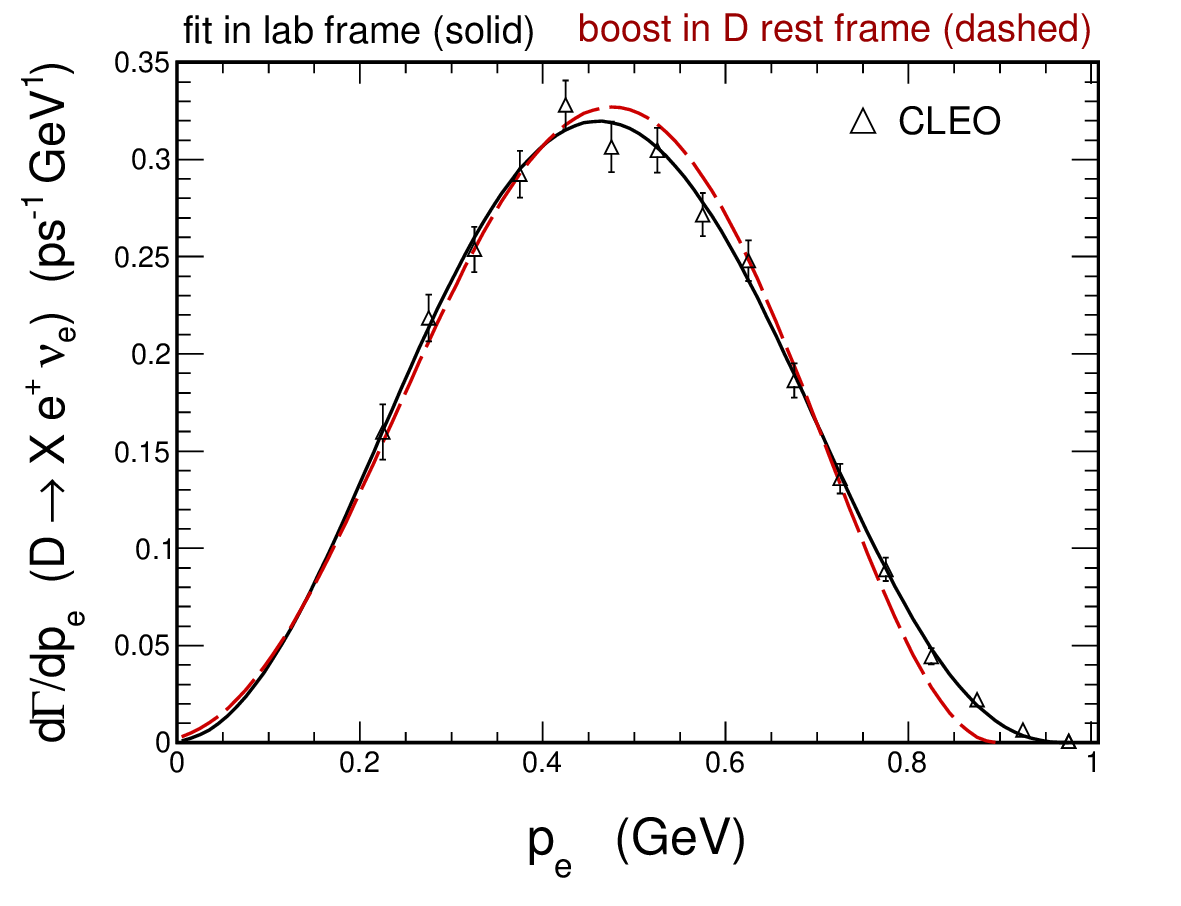}}
\end{minipage}
\begin{minipage}{0.47\textwidth}
 \centerline{\includegraphics[width=1.0\textwidth]{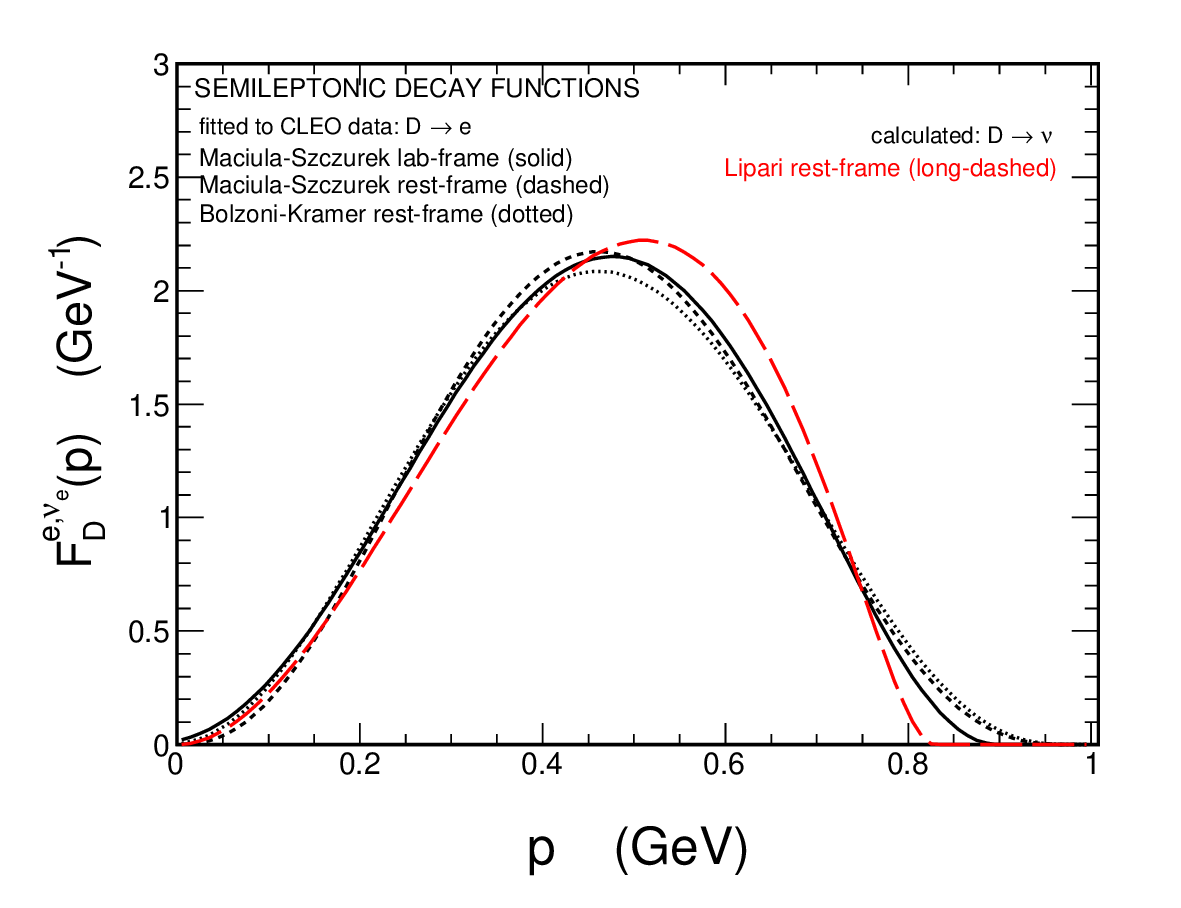}}
\end{minipage}

   \caption{
\small Left panel: Fits to the CLEO data. The solid lines correspond to the parametrizations in the laboratory frames and the dashed lines to the meson rest frames, which represent incorporation of effects related to the non-zero motion of decaying mesons. Right panel: A comparison of different decay functions normalized to $1$ taken from the literature.}
 \label{fig:p-decay-1}
\end{figure}

Both, laboratory and rest frame parametrizations of the semileptonic decay functions for $D$ meson are drawn in the left panel of Fig.~\ref{fig:p-decay-1} together with the CLEO experimental data. Some small differences between the different parametrizations appear only at larger values of electron momentum. The influence of this effect on differential cross sections of leptons is expected to be negligible. As can be seen from the right panel of Fig.~\ref{fig:p-decay-1} our analytical formulae for the decay functions (solid and dashed lines) only slightly differ from the one obtained in a similar approach in Ref.~\cite{Bolzoni:2012kx} (dotted line) and from the one calculated by using the Lipari's formula \cite{Lipari:1993hd} (long-dashed formula) described in Eq.~(\ref{lipari-formula}).

The phenomenological model for production of leptons from the semileptonic $D$-meson decays in hadronic reactions, based on the experimental decay functions described above, has been found to give a very good description of the LHC experimental data collected, e.g. with the ALICE detector \cite{ALICE:2014ivb}.

\subsection{Production of $\nu_{\tau}$ neutrino}

The production mechanism of $\nu_{\tau}$ or ${\bar \nu}_{\tau}$ is a bit
more complicated. The decay of $D_s$ mesons to $\nu_e$ and $\nu_{\mu}$ are often neglected
as the relevant $c \to D_s$ fragmentation fraction is relatively small BR($c \to D_{s}) \approx 8\%$ and further decay branching fractions to $\nu_e$ and $\nu_{\mu}$
are about $2\%$ only. On the other hand, the $D_s$ mesons are also a source of $\nu_{\tau}$ neutrinos/antineutrinos.
The $D_s$ mesons are quite unique in the production of $\nu_{\tau}$,
in particular, decay of $D_s$ mesons is the dominant mechanism of $\nu_{\tau}$ production.

In hadronic reactions such neutrinos/antineutrinos come from
the decay of $D_s$ mesons. There are two mechanisms described shortly
below:
\begin{enumerate}[label=(\alph*)]
\item the direct decay mode: $D_s^+ \to \tau^+ \nu_{\tau}$ with $\mathrm{BR} = 5.32 \pm 0.11 \%$  and
\item the chain decay mode: $D_s^+ \to \tau^+ \to \overline{\nu}_{\tau}  $.
\end{enumerate}
More information can be found in Ref.~\cite{Maciula:2019clb} dedicated to the SHIP fixed-target experiment where the production of the $\nu_{\tau}$ neutrinos in a fixed-target
$p\!+\!^{96}\mathrm{Mo}$ reaction at $\sqrt{s}=27.4$ GeV was discussed. Here we discuss $p+p$ collisions
in their center of mass system which is also laboratory system for the
experimental set up.

\subsubsection{Direct decay of $D_s^{\pm}$ mesons}

The considered here decay channels: $D_s^+ \to \tau^+ \nu_{\tau}$ and
$D_s^- \to \tau^- {\overline \nu}_{\tau}$, which are the sources of 
the direct neutrinos,
are analogous to the standard text book cases of $\pi^+ \to \mu^+
\nu_{\mu}$ and $\pi^- \to \mu^- {\overline \nu}_{\mu}$ decays, discussed
in detail in the past (see e.g. Ref~\cite{Renton:1990td}). The same formalism 
used for the pion decay applies also to the $D_s$ meson decays.
Since pion has spin zero it decays isotropically in its rest frame. 
However, the produced muons are polarized in its direction of motion
which is due to the structure of weak interaction in the Standard
Model. 
The same is true for $D_s^{\pm}$ decays and polarization of
$\tau^{\pm}$ leptons.

As it was explicitly shown in Ref.~\cite{Maciula:2019clb} the $\tau$ lepton takes almost whole energy of the mother $D_s$-meson. This is because of the very similar mass of both particles: $m_{\tau} = 1.777$ GeV and $m_{D_{s} = 1.968}$ GeV. The direct neutrinos take only a small part of the energy and therefore will form the low-energy component of the neutrino flux observed by the FASER$\nu$ experiment.

\subsubsection{Neutrinos from chain decay of $\tau$ leptons}

The $\tau$ decays are rather complicated due to having many possible decay 
channels~\cite{ParticleDataGroup:2022pth}. Nevertheless, all confirmed decays lead to production of 
$\nu_{\tau}$ (${\overline \nu}_{\tau}$). This means total amount 
of neutrinos/antineutrinos produced from $D_s$ decays into $\tau$ lepton 
is equal to the amount of antineutrinos/neutrinos produced in subsequent
$\tau$ decay.
But, their energy distributions will be different \cite{Maciula:2019clb}.

The purely leptonic channels (three-body decays), analogous to the
$\mu^{\pm} \to e^{\pm} ({\overline \nu}_{\mu}/\nu_{\mu})( \nu_e / {\overline \nu}_e)$
decay (discussed e.g. in Refs.~\cite{Renton:1990td,Gaisser:1990vg}) cover only about 35\% of all 
$\tau$ lepton decays. Remaining 65\% are semi-leptonic decays. 
They differ quite drastically from each other and each gives 
slightly different energy distribution for 
$\nu_{\tau}$ (${\overline \nu}_{\tau}$).
In our model for the decay of $D_s$ mesons there
is almost full polarization of $\tau$ particles with respect to 
the direction of their motion.

Since $P_{\tau^+} = -P_{\tau^-}$ and 
the angular distributions of polarized $\tau^{\pm}$ are
antisymmetric with respect to the spin axis the resulting distributions 
of $\nu_{\tau}$ and ${\overline \nu}_{\tau}$
from decays of $D_s^{\pm}$ are then identical, consistent with CP 
symmetry (see e.g. Ref.~\cite{Barr:1988rb}).

In the numerical calculations of $\nu_{\tau}$ neutrinos/antineutrinos we use a sample of
10$^5$ decays generated before by the dedicated TAUOLA 
program \cite{TAUOLA}
in the $D_s$ center of mass. The distributions (event by event)
are transformed then to the proton-proton center of mass system
which is also laboratory system where the measurement take place.
Then the momentum/rapidity distributions are obtained and cuts
on (pseudo)rapidity of neutrinos are imposed.

We are interested in energy distribution (fluxes) of different
kinds of neutrinos/antineutrinos. In the present paper we do not simulate interactions of neutrinos
with a dedicated target so we will not estimate actual number of 
experimental events for FASER$\nu$. Such number would depend on 
a given target used in the experiment. As discussed in the result section
already the flux of neutrinos/anineutrinos corresponding
to different mechanisms will allow to draw very interesting conclusions,
especially on the intrinsic charm in the proton.

\section{Numerical results}

\begin{figure}[!h]
\begin{minipage}{0.45\textwidth}
  \centerline{\includegraphics[width=1.0\textwidth]{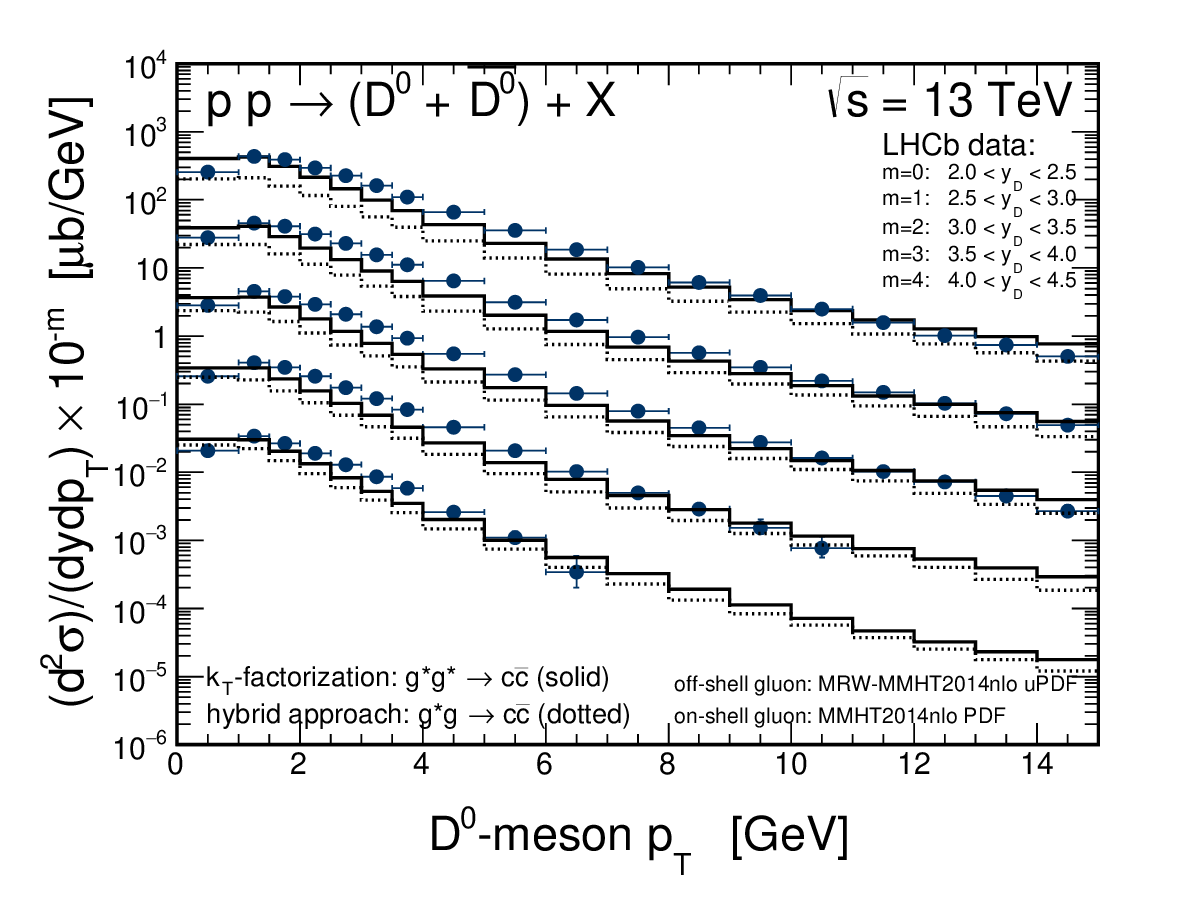}}
\end{minipage}
\begin{minipage}{0.45\textwidth}
  \centerline{\includegraphics[width=1.0\textwidth]{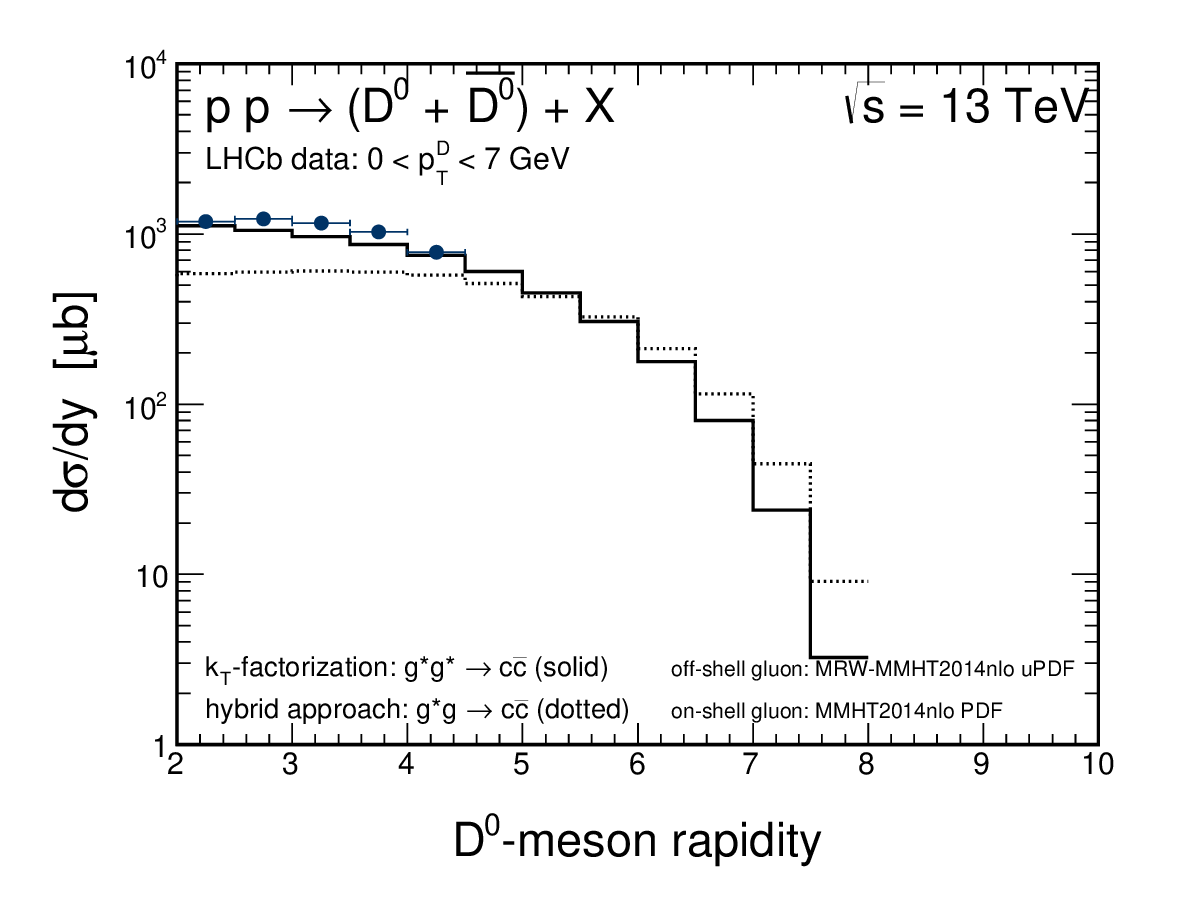}}
\end{minipage}
  \caption{
\small Transverse momentum distirbutions
for different windows of rapidity (left panel) and rapidity distirbution (right panel) of $D^0 + {\bar D}^0$ mesons at $\sqrt{s} = 13$ TeV, obtained with the MRW-MMHT2014nlo gluon uPDF together with the LHCb data \cite{LHCb:2015swx}.
Details are specified in the figure.
}
\label{fig:LHCb_KMR}
\end{figure}

We start our presentation of results from the forward production 
of $D$ mesons at the LHC energy $\sqrt{s} = 13$ TeV within the LHCb experiment rapidity acceptance, i.e. $2 < y < 4.5$.
In Fig.~\ref{fig:LHCb_KMR} we show transverse
momentum and rapidity distributions of $D^0 + {\bar D}^0$ calculated within the full $k_{T}$-factorization approach (solid histograms) as well as within the hybrid model (dotted histograms), for the MRW-MMHT2014nlo gluon uPDFs. Here and in the following the on-shell collinear parton distributions are taken from the MMHT2014nlo PDF set \cite{Harland-Lang:2014zoa}. The theoretical predictions are compared to the LHCb experimental data \cite{LHCb:2015swx}.
Here, a very good agreement with the LHCb data is obtained with the full $k_{T}$-factorization calculations. The hybrid model seems to underestimate the experimental distributions at more central rapidities, however, both predictions starts to coincide in more forward region, i.e. $4.5 < y < 6.5$, beyond the LHCb detector coverage. In the far-forward region ($y > 6.5$) the hybrid approach leads to slightly larger cross sections than the full $k_{T}$-factorization.

\begin{figure}[!h]
\begin{minipage}{0.45\textwidth}
  \centerline{\includegraphics[width=1.0\textwidth]{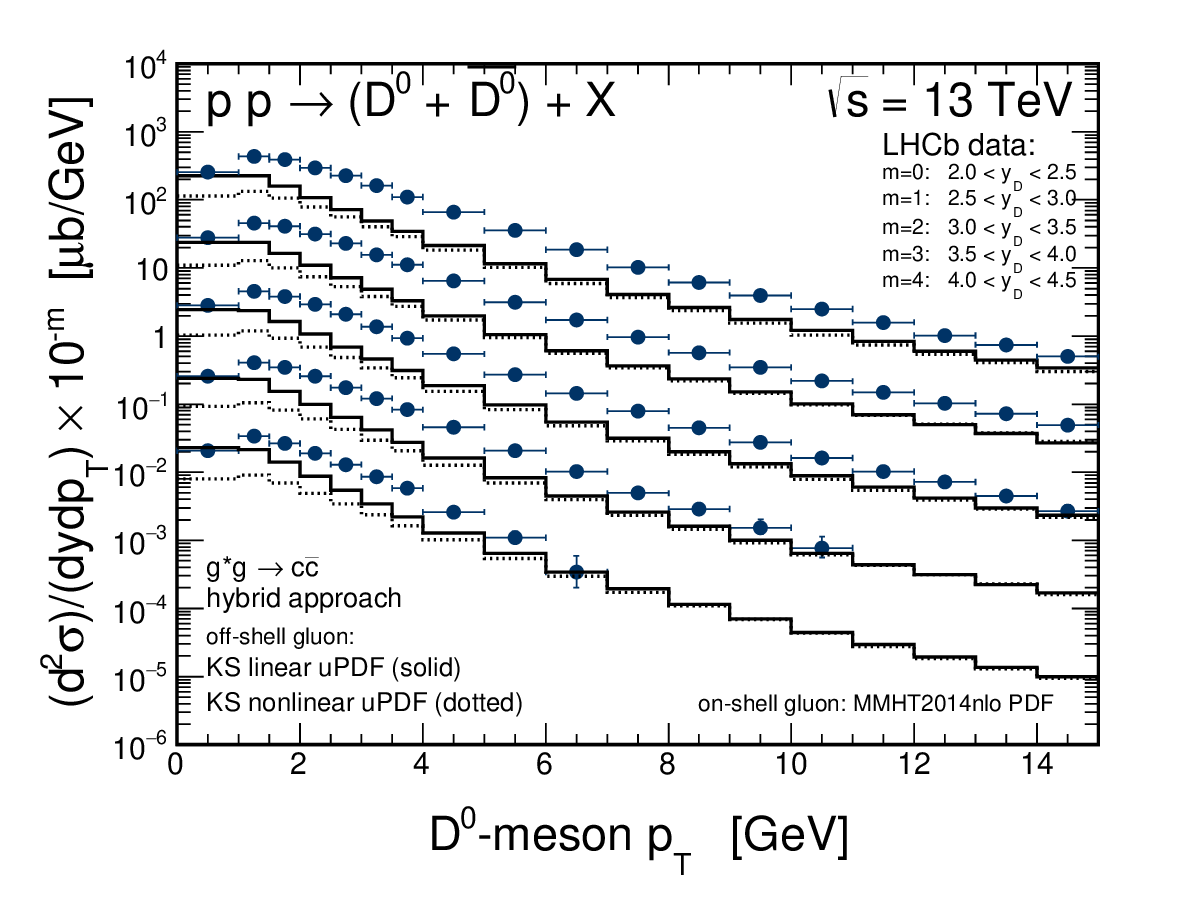}}
\end{minipage}
\begin{minipage}{0.45\textwidth}
  \centerline{\includegraphics[width=1.0\textwidth]{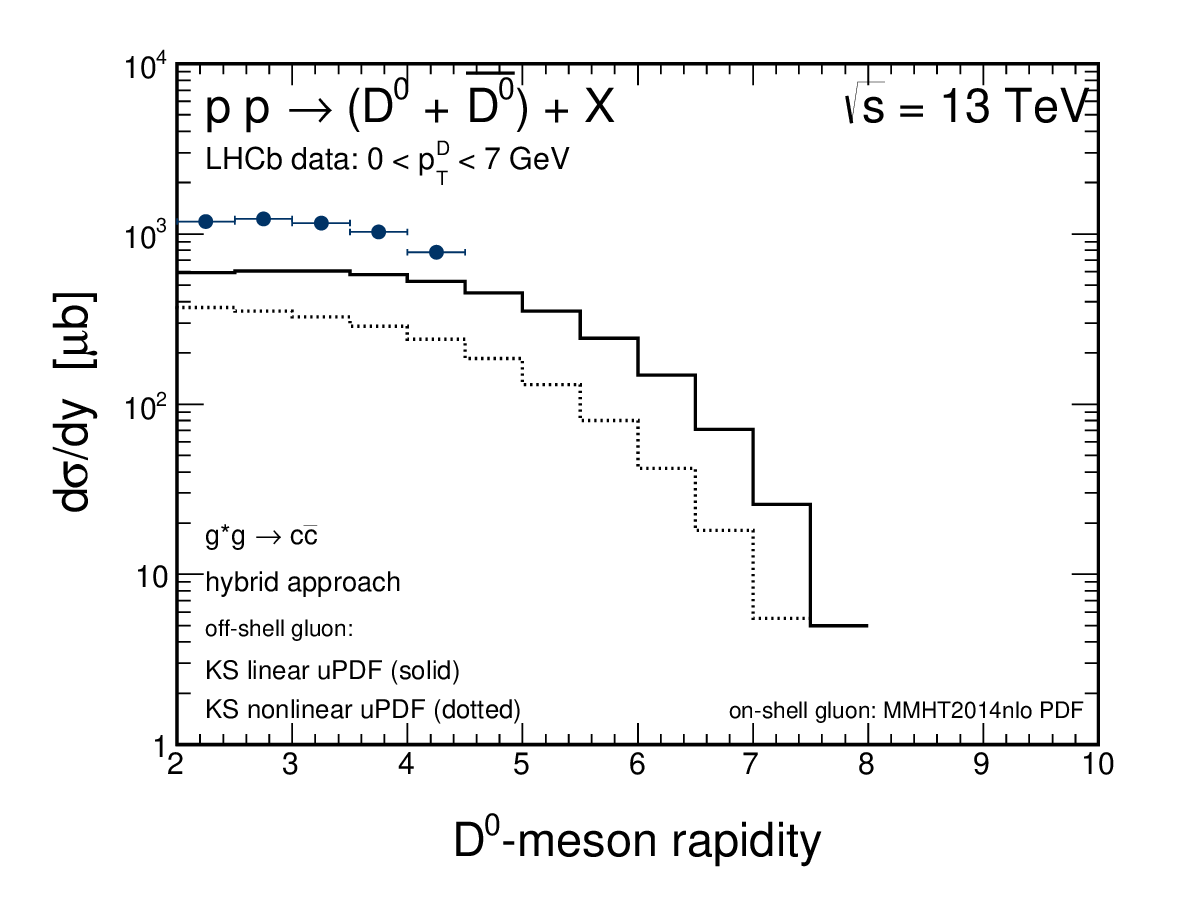}}
\end{minipage}
  \caption{
\small Transverse momentum distirbutions
for different windows of rapidity (left panel) and rapidity distirbution (right panel) of $D^0 + {\bar D}^0$ mesons at $\sqrt{s} = 13$ TeV, obtained in the hybrid approach with the KS linear and KS nonlinear gluon uPDFs together with the LHCb data \cite{LHCb:2015swx}.
Details are specified in the figure.
}
\label{fig:LHCb_KS}
\end{figure}

In Fig.~\ref{fig:LHCb_KS} we show a similar theory-to-data comparison as above, but here we plot numerical results obtained with the KS linear (solid histograms) and nonlinear (dotted histograms) gluon uPDFs. Both calculations here are obtained within the hybrid approach. The KS uPDFs are available only for $x < 10^{-2}$ so they cannot be used on the large-$x$ side in the full $k_{T}$-factorization calculations, especially in the case of forward charm production. As we can see the difference between predictions of the linear and nonlinear uPDFs appear only at very small transverse momenta. Unfortunately, both of them visibly underestimate the LHCb data points, hovewer, the discrepancy
seems to decrease when moving to more forward rapidities. Therefore, one should not discard them in the far-forward limit.

\begin{figure}[!h]
\begin{minipage}{0.3\textwidth}
  \centerline{\includegraphics[width=1.0\textwidth]{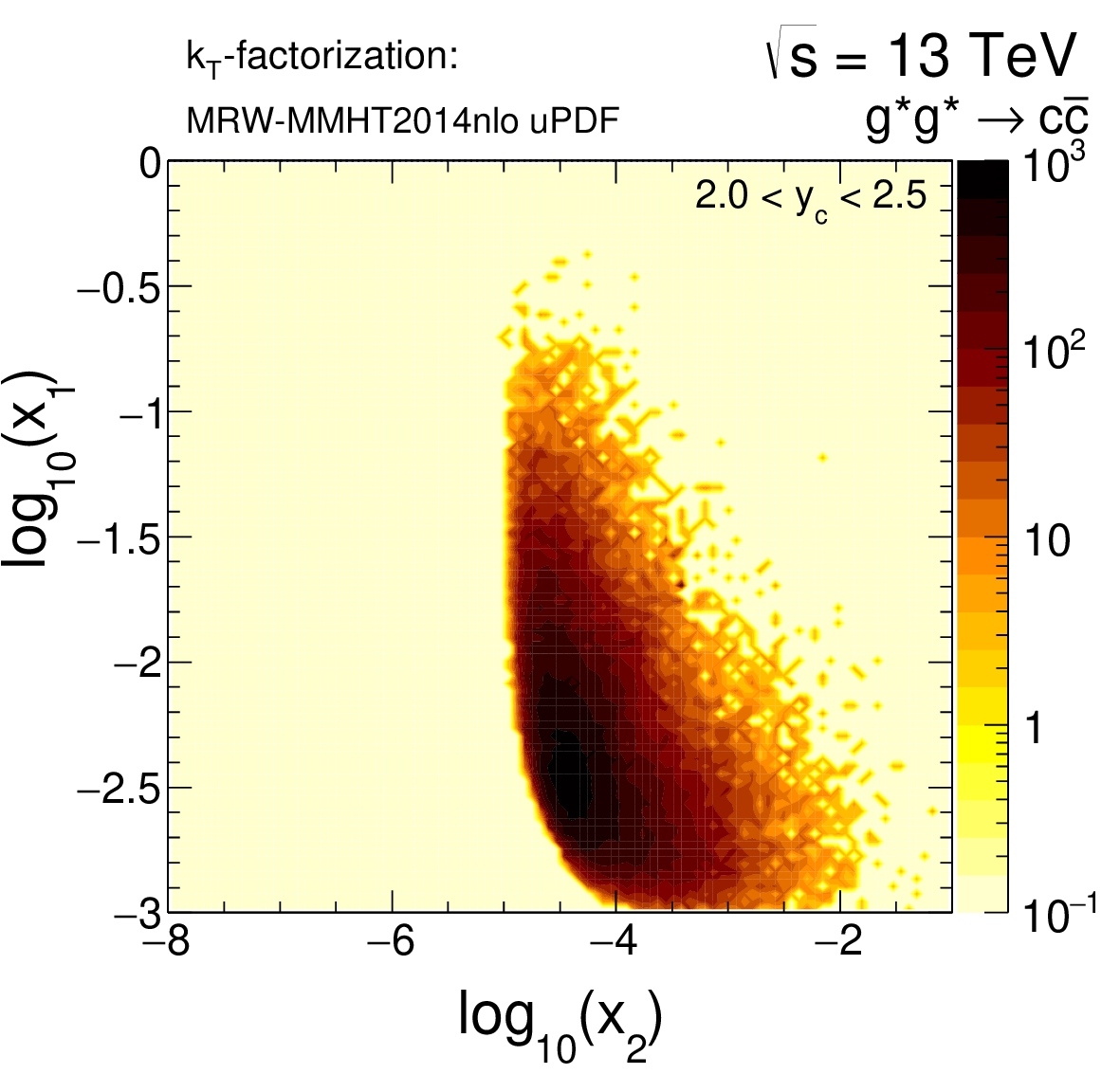}}
\end{minipage}
\begin{minipage}{0.3\textwidth}
  \centerline{\includegraphics[width=1.0\textwidth]{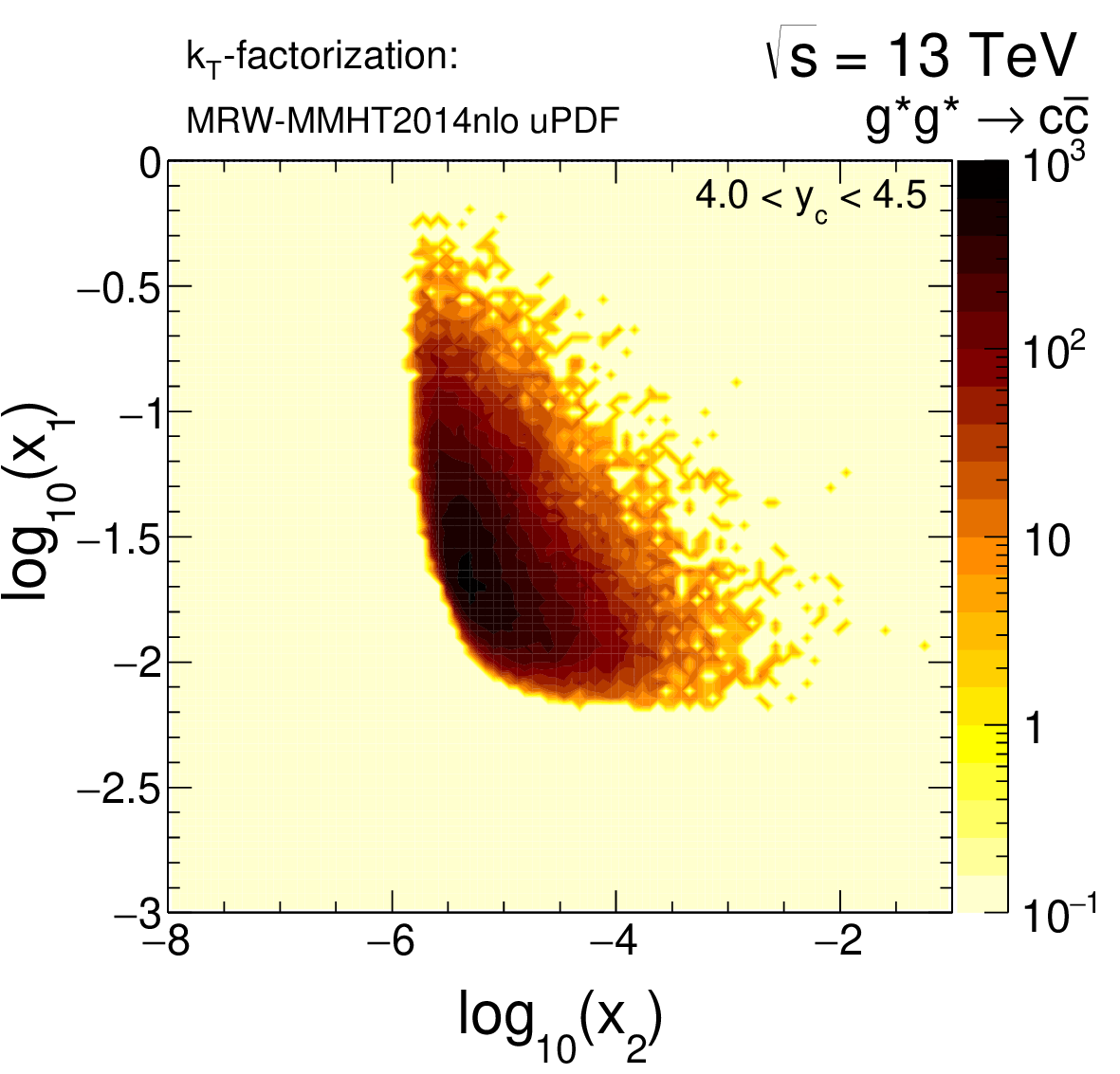}}
\end{minipage}
\begin{minipage}{0.3\textwidth}
  \centerline{\includegraphics[width=1.0\textwidth]{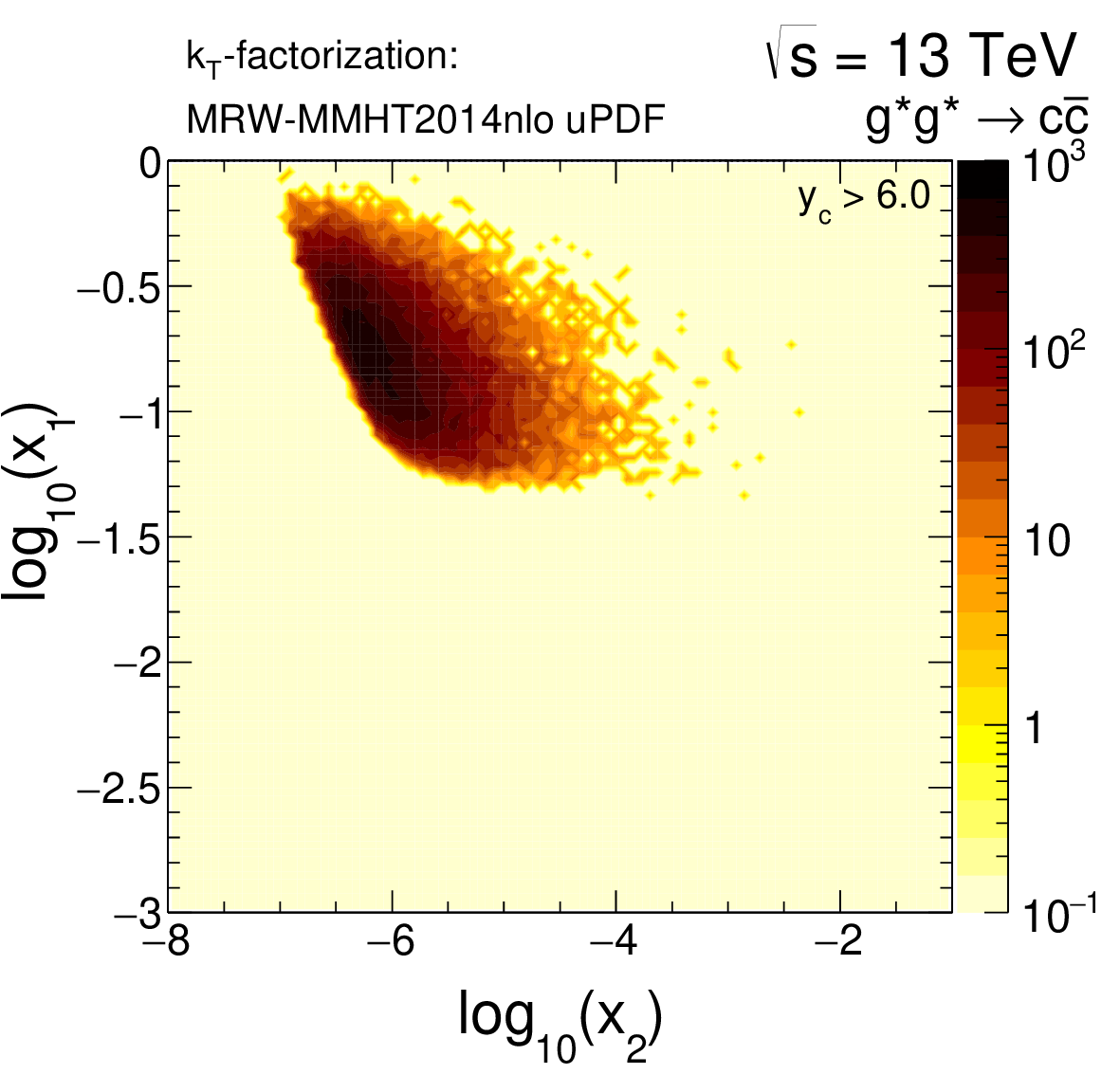}}
\end{minipage}
  \caption{
\small Two-dimensional distribution in $\log_{10}(x_1)$ and  $\log_{10}(x_2)$
for different windows of rapidity calculated in the full $k_{T}$-factorization approach for the MRW-MMHT2014nlo uPDF.
}
\label{fig:log10x1log10x2_KMR}
\end{figure}

\begin{figure}[!h]
\begin{minipage}{0.3\textwidth}
  \centerline{\includegraphics[width=1.0\textwidth]{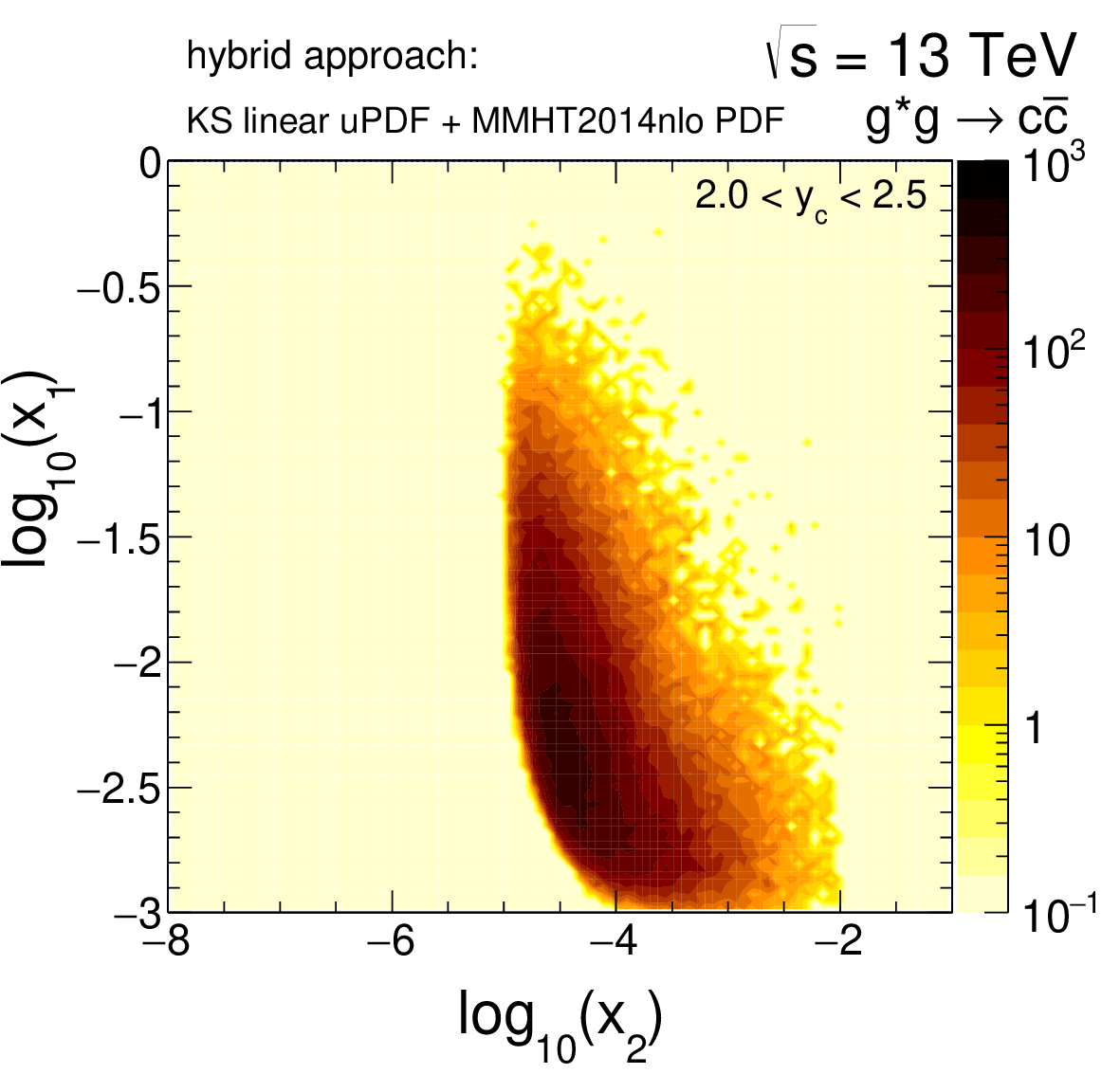}}
\end{minipage}
\begin{minipage}{0.3\textwidth}
  \centerline{\includegraphics[width=1.0\textwidth]{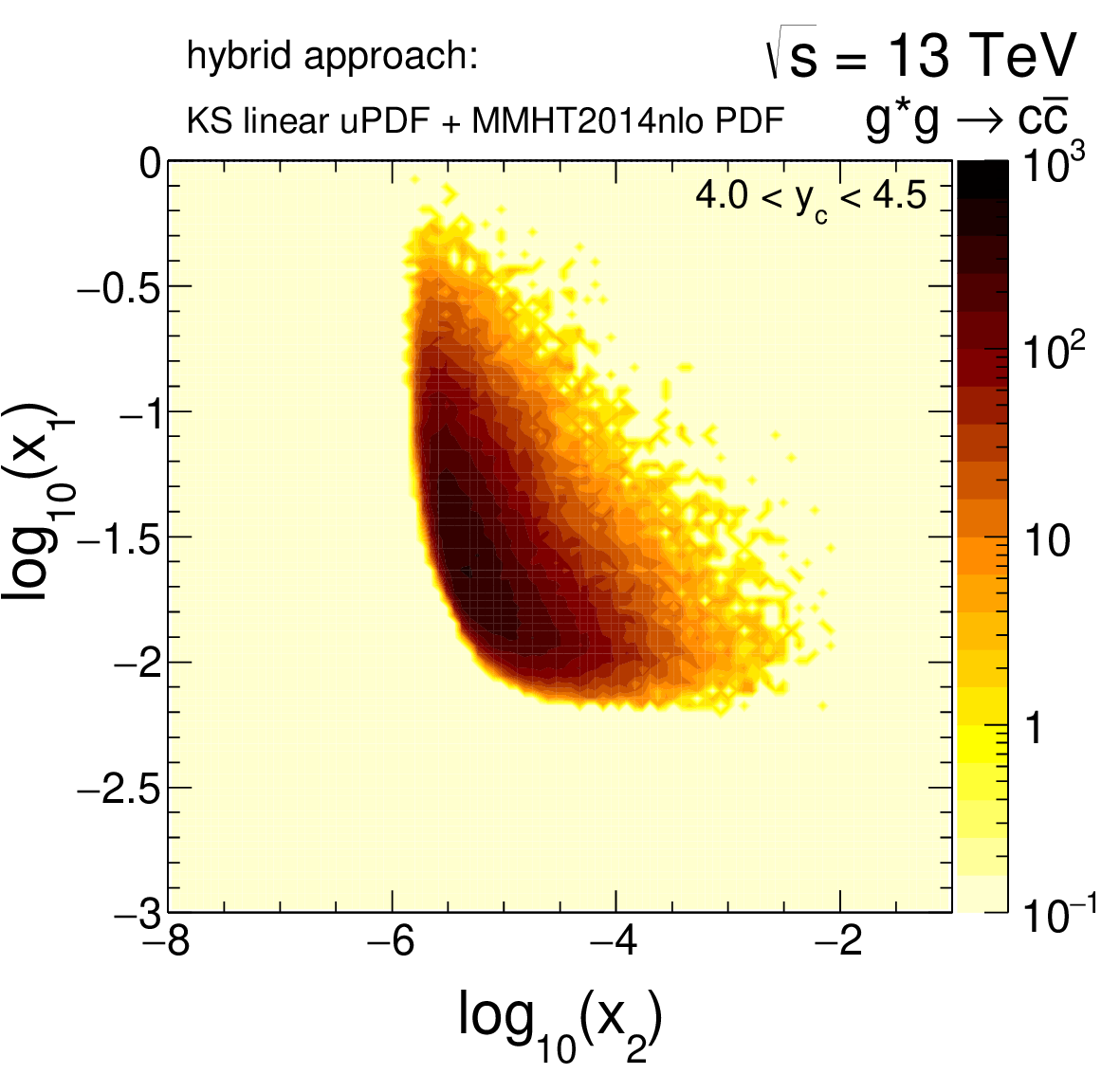}}
\end{minipage}
\begin{minipage}{0.3\textwidth}
  \centerline{\includegraphics[width=1.0\textwidth]{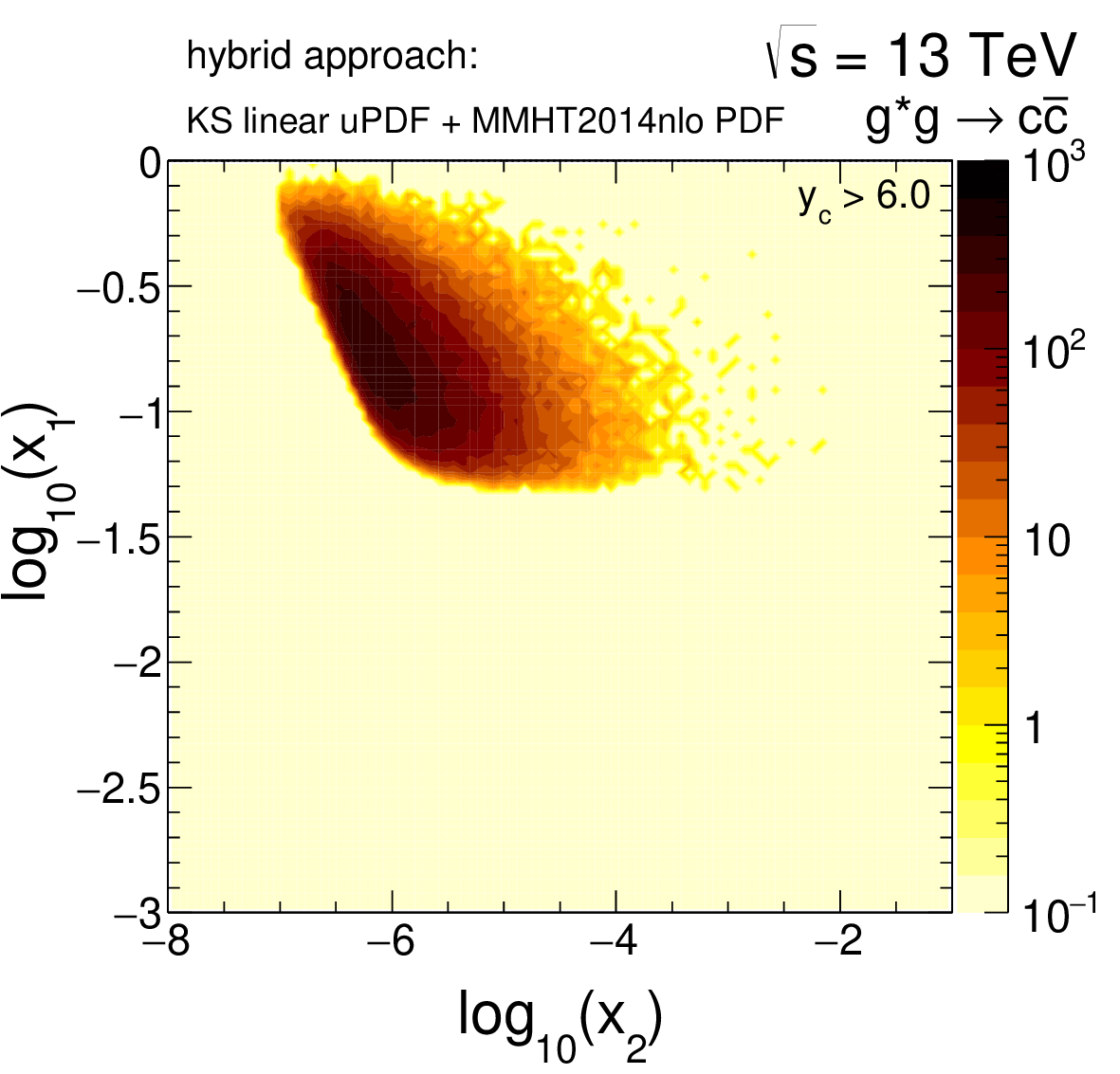}}
\end{minipage}
  \caption{
\small Two-dimensional distribution in $\log_{10}(x_1)$ and  $\log_{10}(x_2)$
for different windows of rapidity calculated in the hybrid model for the KS linear uPDF.
}
\label{fig:log10x1log10x2_KS}
\end{figure}

In Figs.~\ref{fig:log10x1log10x2_KMR} and ~\ref{fig:log10x1log10x2_KS} we show the region of longitudinal
momentum fractions of gluons entering the fusion process for different
windows of rapidity. We observe that even in the current LHCb acceptance one deals with the very asymmetric configurations where $x_1 \gg x_2$.
The situation depends on the rapidity interval and for most forward LHCb
interval of rapidity one probes $x_2$ down to $\approx 10^{-5}$,
(a typical region where one could expect the onset of saturation
effects) and simultaneously $x_1$ above $10^{-2}$. The kinematical configuration becomes
even more interesting and challenging when approaching the far-forward region, taking e.g. $y > 6.0$, where 
one could probe $x_{1} > 0.1$ and $x_{2} < 10^{-6}$.

\begin{figure}[!h]
\begin{minipage}{0.45\textwidth}
  \centerline{\includegraphics[width=1.0\textwidth]{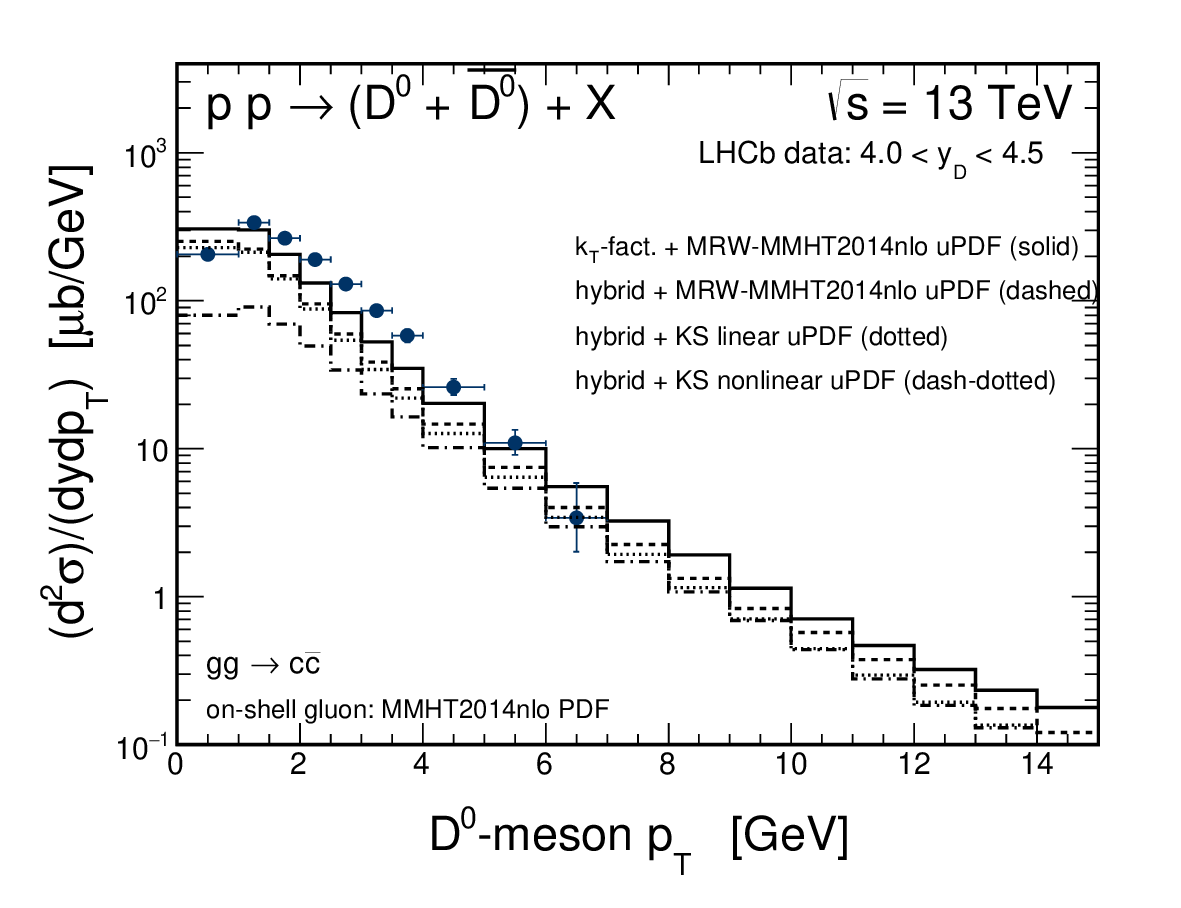}}
\end{minipage}
\begin{minipage}{0.45\textwidth}
  \centerline{\includegraphics[width=1.0\textwidth]{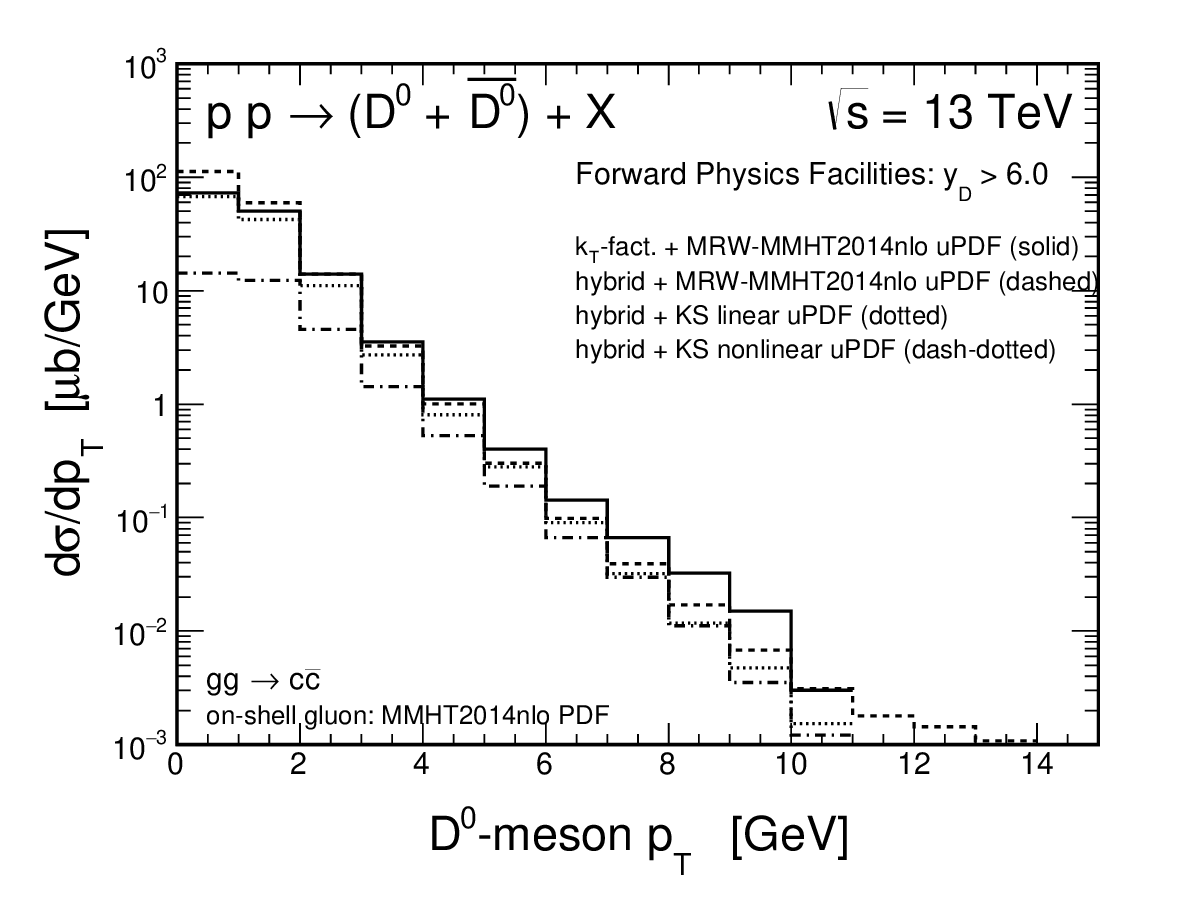}}
\end{minipage}
  \caption{
\small Transverse momentum distribution of $D^0 + {\bar D}^0$ mesons
for 4.0 $< y <$ 4.5 (left panel) and $y >$ 6 (right panel)
for $\sqrt{s}$ = 13 TeV.
We show result for the $k_t$-factorization as well as for
the hybrid approach.
}
\label{fig:4to4.5andgt6}
\end{figure}

Let us concentrate now on the most forward $D$ meson production.
In the left panel of Fig.~\ref{fig:4to4.5andgt6} we show result 
for the most forward LHCb rapidity bin (4 $< y <$ 4.5) obtained
within the $k_T$-factorization approach as well as results for 
the hybrid approach. The hybrid approach for the MRW-MMHT2014nlo and the KS linear uPDF (dashed and dotted lines, respectively) gives only somewhat smaller cross section than the $k_T$-factorization approach for the MRW-MMHT2014nlo uPDF (solid lines). Only the hybrid predictions for the KS nonlinear uPDF (dash-dotted lines)
seems to be completely disfavoured by the LHCb data.

In the right panel we show similar results but for
$y >$ 6, not accesible so far at the LHC for the $D$ meson production. Here, the prediction of the full $k_{T}$-factorization for the MRW-MMHT2014nlo uPDF coincide with those calculated with the hybrid model for the MRW-MMHT2014nlo and the KS linear uPDF. The three different calculations lead to a very similar results in the far-forward limit of charm production, what leaves a slight freedom of their choice.

\begin{figure}[!h]
\begin{minipage}{0.45\textwidth}
  \centerline{\includegraphics[width=1.0\textwidth]{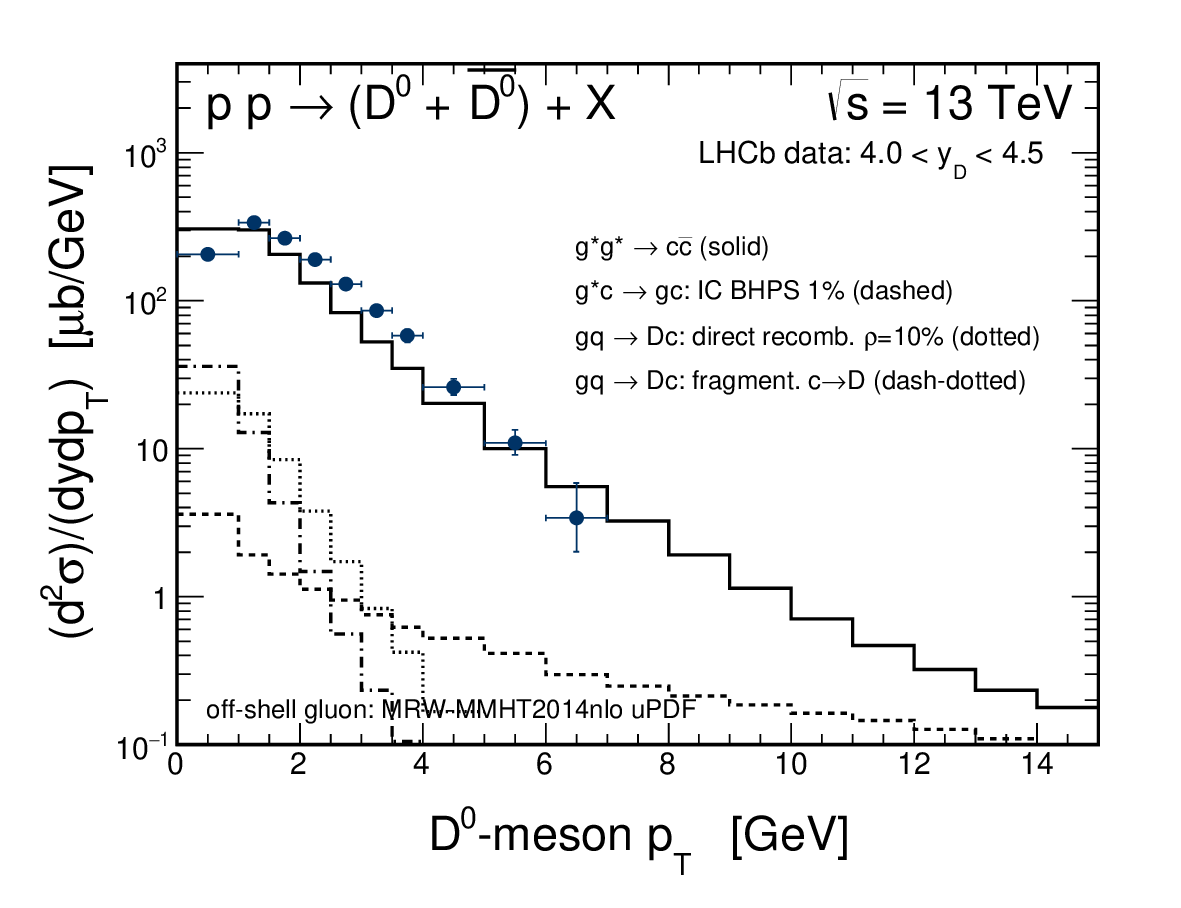}}
\end{minipage}
\begin{minipage}{0.45\textwidth}
  \centerline{\includegraphics[width=1.0\textwidth]{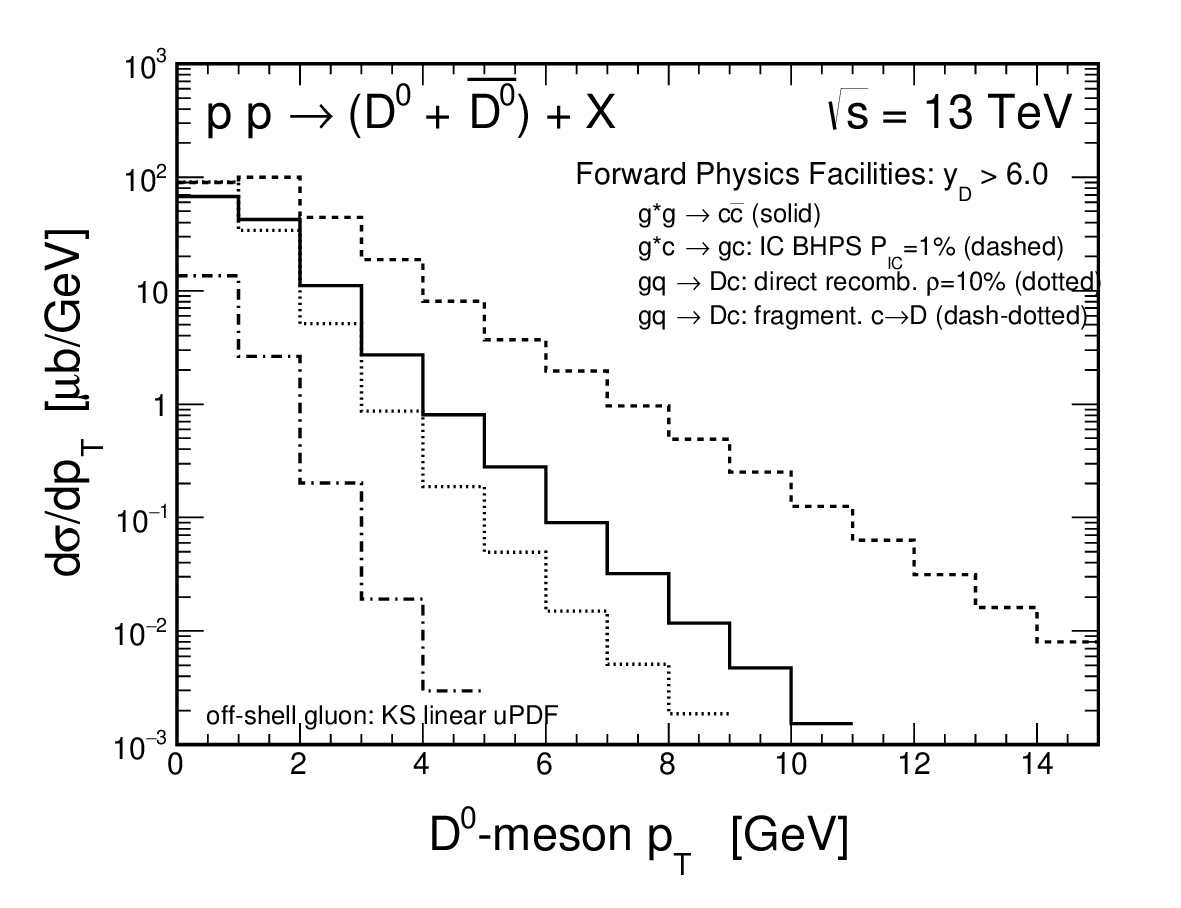}}
\end{minipage}
  \caption{
\small Transverse momentum distributions of $D^0 + {\bar D}^0$ for the
most forward LHCb rapidity bin (left panel) and for $y >$ 6 (right
panel) for $\sqrt{s}$ = 13 TeV. We show separately the $gg$-fusion (solid), the intrinsic charm (dashed) and the two contributions of recombination (dotted and dash-dotted) mechanisms.
}
\label{fig:subleading_pt}
\end{figure}

So far we have considered only the dominant at midrapidity
gluon-gluon fusion mechanism of charm production. Now we shall consider the subdominant at
midrapidities the intrinsic charm and the recombination contributions.
The free parameters: $P_{IC}$ for the intrinsic charm and $\rho$ for the recombination
mechanisms used in this calculation are those found in our recent 
analyses of the high-energy neutrino IceCube data \cite{Goncalves:2021yvw} and of the fixed-target LHCb charm data \cite{Maciula:2022otw}.
In Fig.~\ref{fig:subleading_pt} we show transverse momentum
distributions for the most forward measured rapidity bin
of the LHCb collaboration (left panel) and for $y >$ 6 (right panel).
The contributions of the subleading mechanisms in the
region measured by the LHCb seem completely negligible. The situation changes dramatically for larger
rapidities. There the intrinsic charm contribution (dashed lines) becomes larger than, and the recombination contribution (dotted and dash-dotted lines)
is of a similar size as, the well known standard $gg$-fusion component calculated here in the hybrid model (solid lines).

\begin{figure}[!h]
\begin{minipage}{0.45\textwidth}
  \centerline{\includegraphics[width=1.0\textwidth]{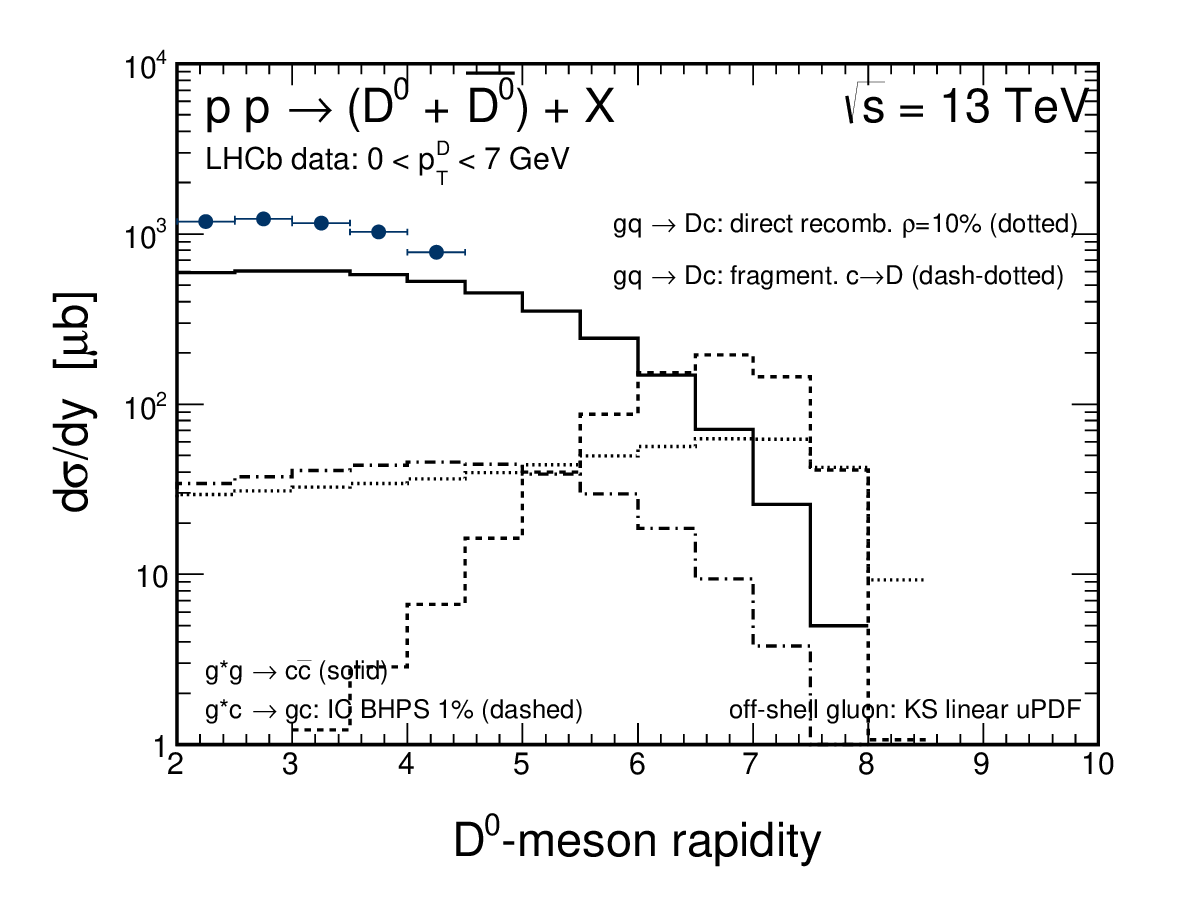}}
\end{minipage}
  \caption{
\small Rapidity distribution of $D^0 + {\bar D}^0$ for $\sqrt{s} =$ 13
TeV. We show separately the $gg$-fusion (solid), the intrinsic charm (dashed) and the two contributions of recombination (dotted and dash-dotted) mechanisms.
}
\label{fig:subleading_y}
\end{figure}

The rapidity distribution of $D^0 + {\bar D}^0$, shown in Fig.~\ref{fig:subleading_y}, nicely summarizes
the situation. Both the IC and the recombination contributions have a distinct maximum
at $y \sim$ 7, where they dominate over the standard mechanism. In the far-forward region the IC contribution is larger than the recombination and the standard components approximately by a factor of $3$ and $6$, respectively. The relative contribution of the IC to the forward charm cross section could be even larger if a cut on low meson transverse momenta is imposed.

\begin{figure}[!h] 
\begin{minipage}{0.45\textwidth}
  \centerline{\includegraphics[width=1.0\textwidth]{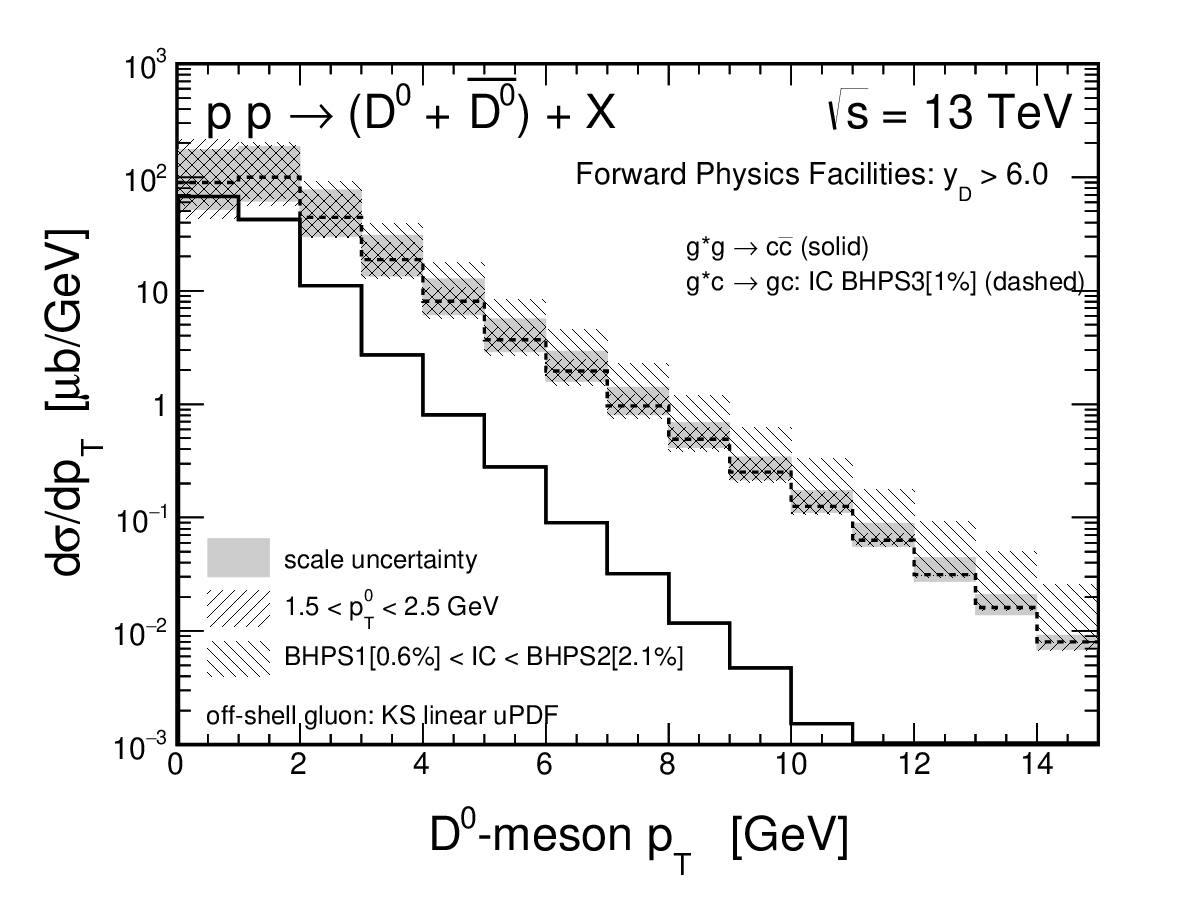}}
\end{minipage}
\begin{minipage}{0.45\textwidth}
  \centerline{\includegraphics[width=1.0\textwidth]{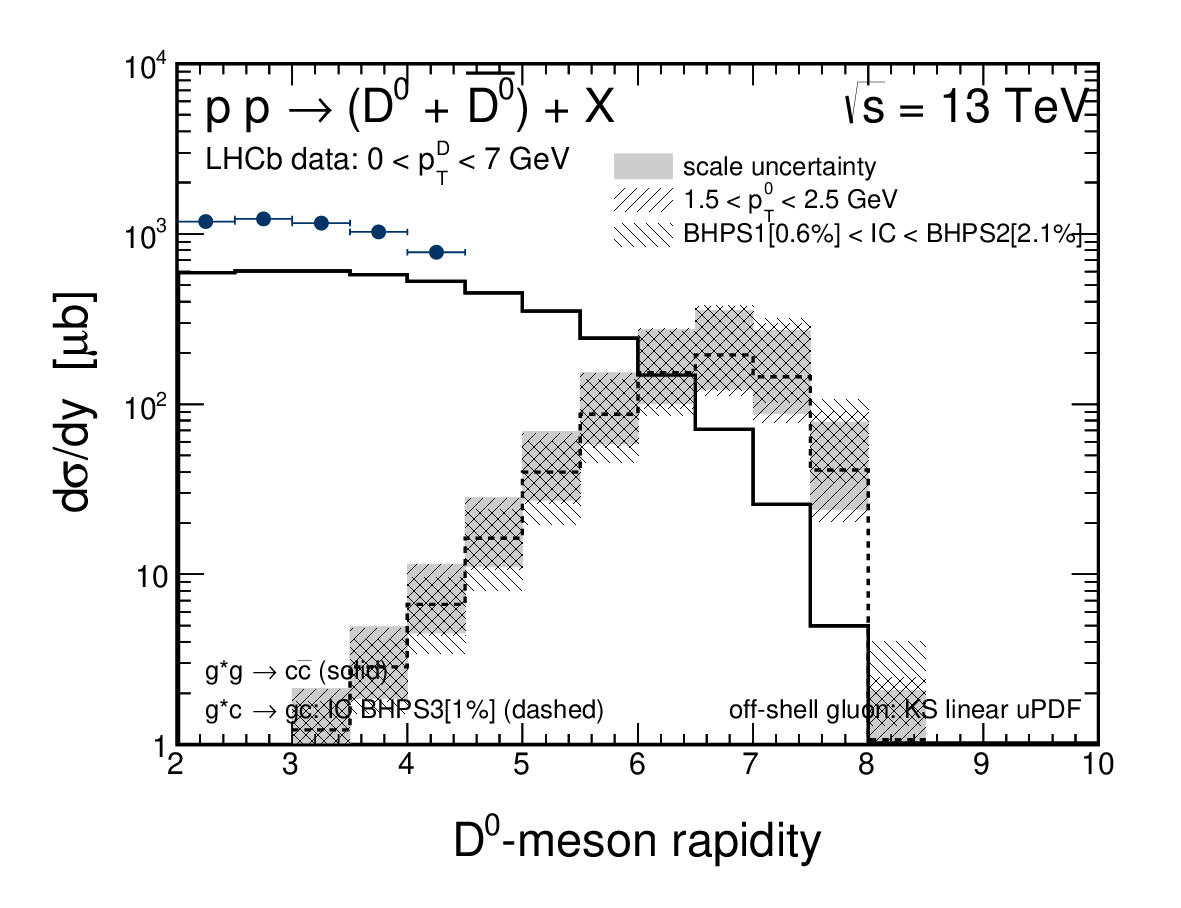}}
\end{minipage}
  \caption{
\small The scale, $P_{IC}$ and $p_{T0}$ uncertainty bands for the IC contribution.
}
\label{fig:IC_uncertainty}
\end{figure}

\begin{figure}[!h]
\begin{minipage}{0.45\textwidth}
  \centerline{\includegraphics[width=1.0\textwidth]{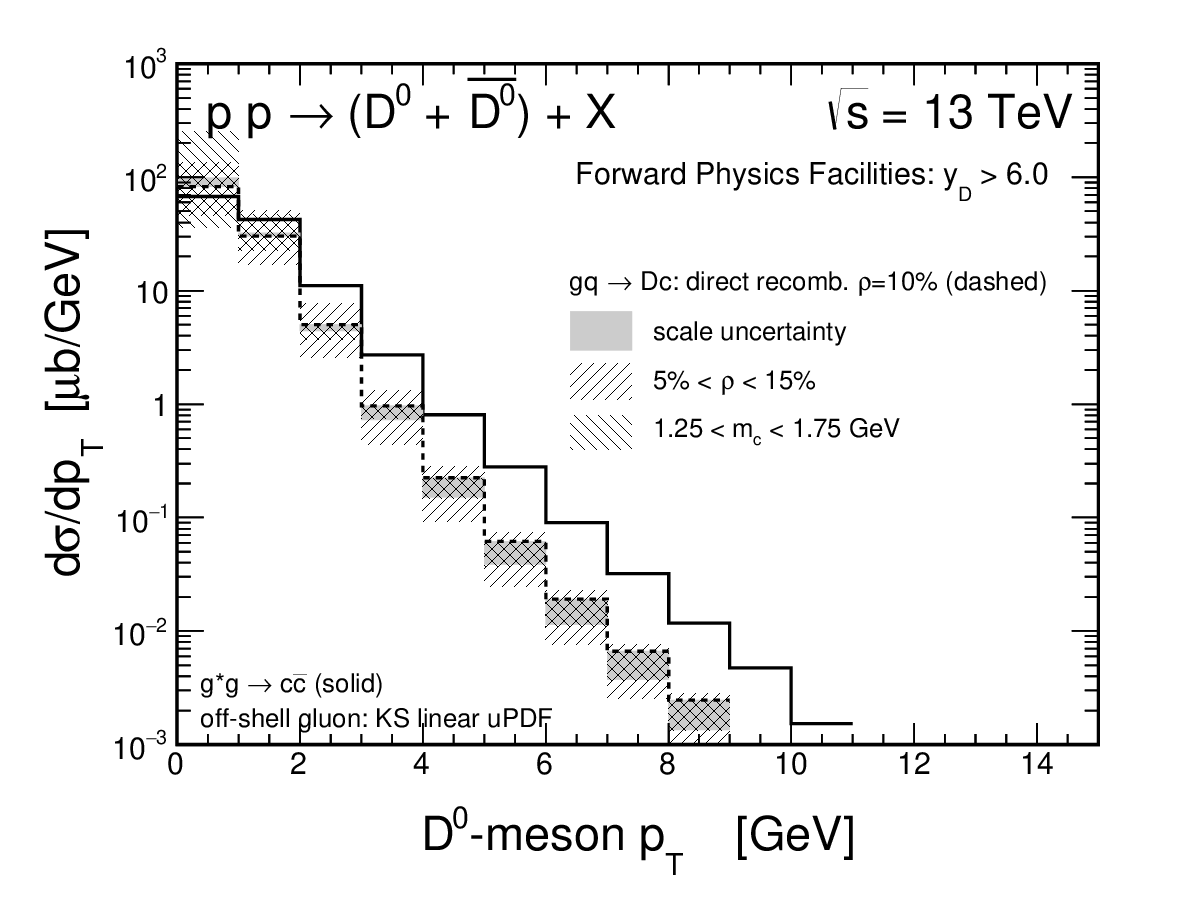}}
\end{minipage}
\begin{minipage}{0.45\textwidth}
  \centerline{\includegraphics[width=1.0\textwidth]{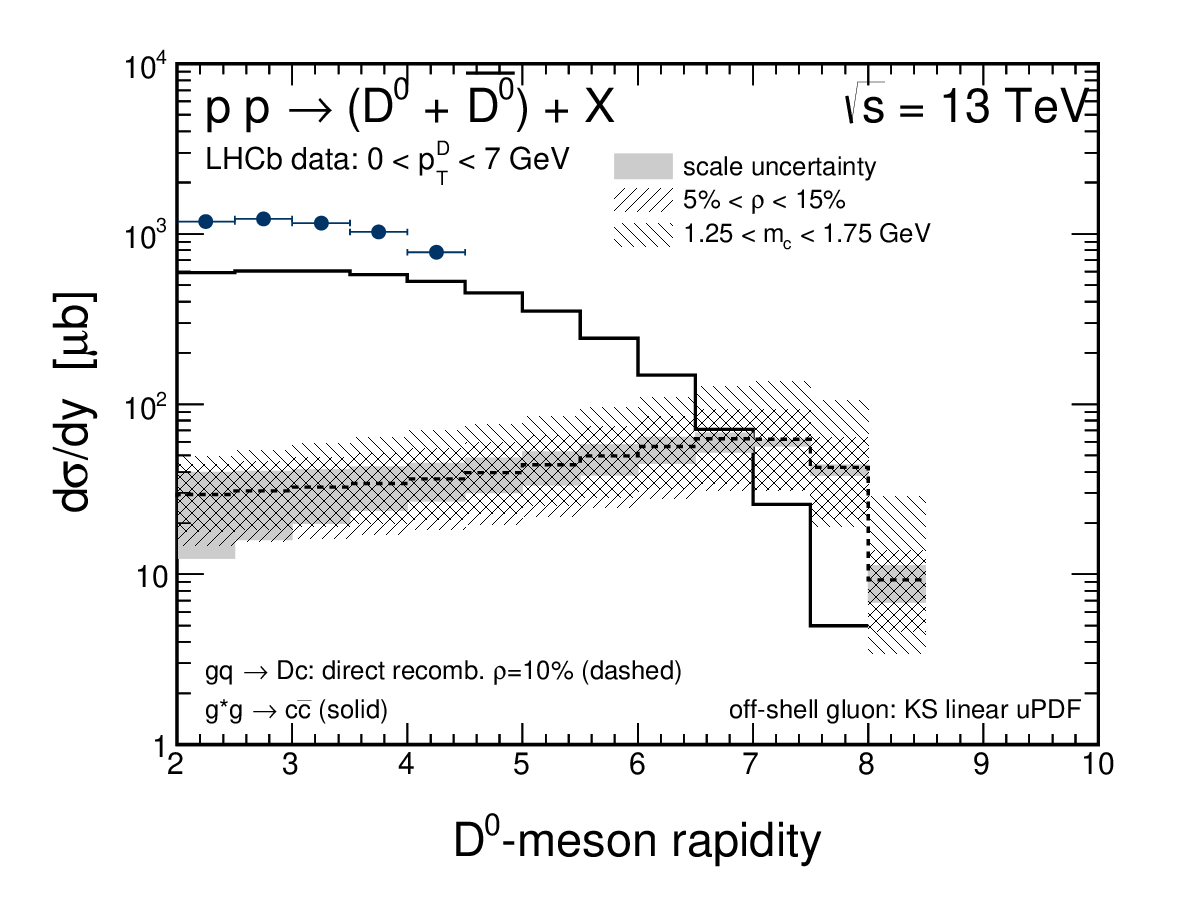}}
\end{minipage}
  \caption{
\small The scale, $\rho$ and $m_{c}$ uncertainty bands for the direct recombination contribution.
}
\label{fig:RECOMB_uncertainty}
\end{figure}

The uncertainties of the calculations of the standard charm production mechanism have been discussed many times in our previous studies (see e.g. Ref.~\cite{Maciula:2013wg}) and will be not repeated here. How uncertain are the IC and the recombination contributions is shown in
Figs.~\ref{fig:IC_uncertainty} and \ref{fig:RECOMB_uncertainty}, respectively. In the case of the intrinsic charm mechanism we plot uncertainties 
related to the scales, the $p_{T0}$ parameter, as well as due to the $P_{IC}$ probability in the BHPS model. In the case of the recombination component we show uncertainties related to the scales, charm quark mass and the recombination probability $\rho$. The uncertainties are quite sizeable, however, even when taking the lower limits of the IC and the recombination predictions one could expect a visible enhancement of the forward charm cross section with respect to the standard calculations. The scenario with upper limits can be examined by the future FPF LHC data.

\begin{figure}[!h]
\begin{minipage}{0.45\textwidth}
  \centerline{\includegraphics[width=1.0\textwidth]{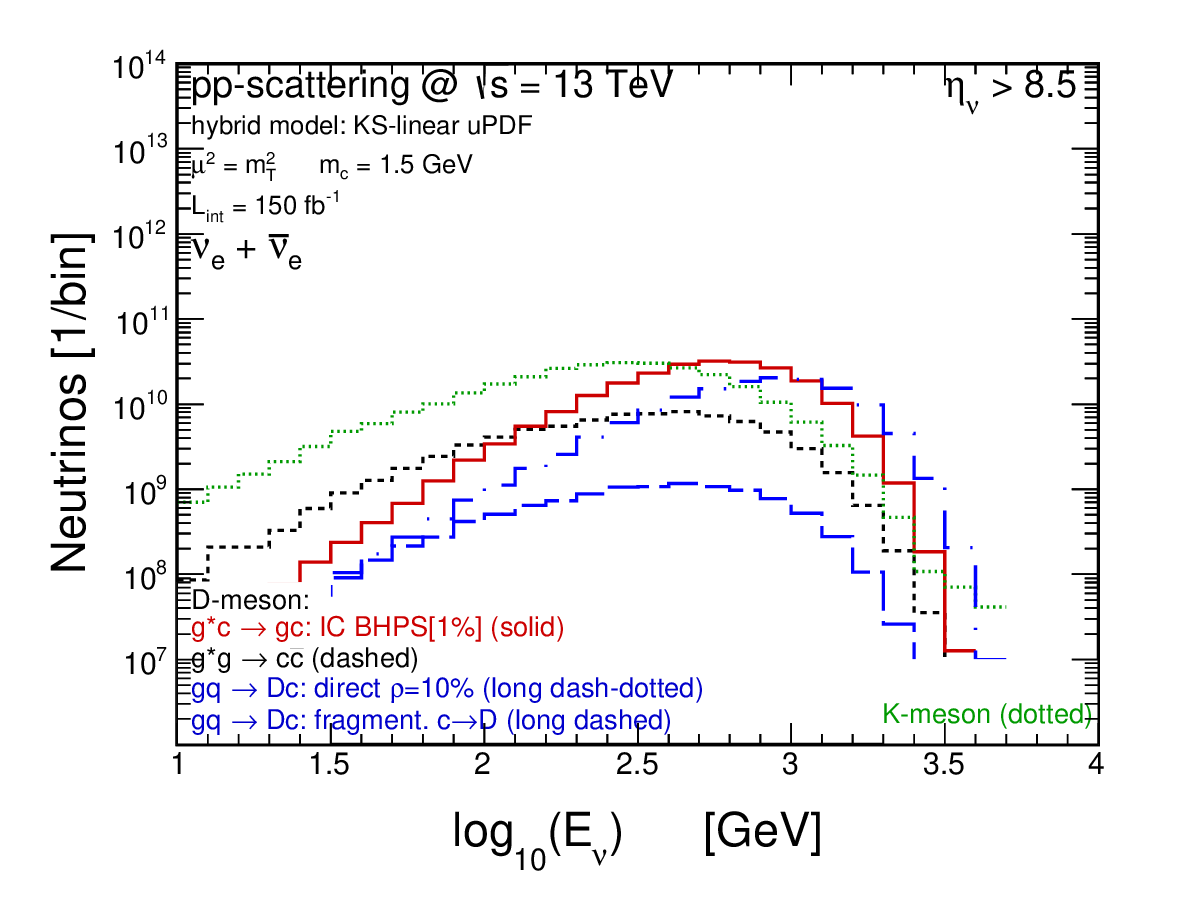}}
\end{minipage}
  \caption{
\small Energy distribution of electron neutrinos+anineutrinos for 
$\eta >$ 8.5 (FASER$\nu$).
}
\label{fig:electron_neutrinos}
\end{figure}

Now we proceed to neutrino/antineutrino production.
In Fig.~\ref{fig:electron_neutrinos} we show energy distribution
of $\nu_e + {\bar \nu}_e$ calculated for the $\sqrt{s} = 13$ TeV including the designed psuedorapidity acceptance $\eta > 8.5$ of the FASER$\nu$ experiment. Here and in the following, the numbers of neutrinos is obtained for the integrated luminosity $L_{\mathrm{int}}= 150  \;\mathrm{fb}^{-1}$. In addition to the production from the
semileptonic decays of $D$ mesons we show contribution from the decay
of kaons (dotted line) taken from \cite{Kling:2021gos}.
The gluon-gluon fusion contribution is quite small, visibly smaller than the kaon contribution.
Both the IC and recombination contributions may be seen as an
enhancement over the contribution due to 
conventional kaons in the neutrino energy distribution at neutrino energies $E_{\nu} > 1$ TeV, however, size of the effect is rather small. An identification of 
the subleading contributions will require a detailed comparison to 
the FASER$\nu$ data. Here the recombination and IC contributions
may be of a similar order.

The situation for muon neutrinos is much more difficult as here
a large conventional contribution from charged pion decays enters \cite{Kling:2021gos}.
Here the IC and recombination contributions are covered by
the $\pi \to \nu_{\mu}$ (dot-dot-dashed), $K \to \nu_{\mu}$ (dotted)
contributions even at large neutrino energies.

\begin{figure}[!h]
\begin{minipage}{0.45\textwidth}
  \centerline{\includegraphics[width=1.0\textwidth]{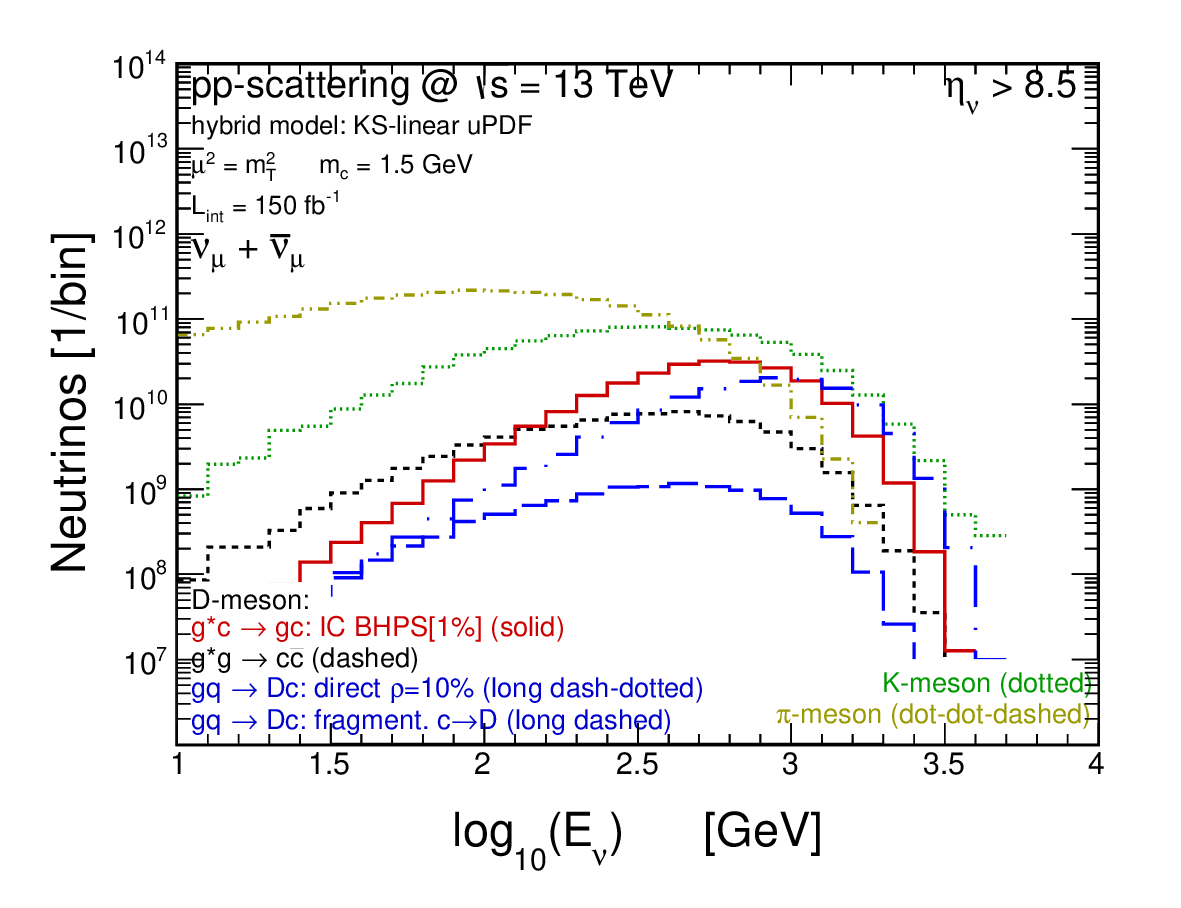}}
\end{minipage}
  \caption{
\small Energy distribution of muon neutrinos+anineutrinos for 
$\eta >$ 8.5 (FASER$\nu$).
}
\label{fig:muon_neutrinos}
\end{figure}

\begin{figure}[!h]
\begin{minipage}{0.45\textwidth}
  \centerline{\includegraphics[width=1.0\textwidth]{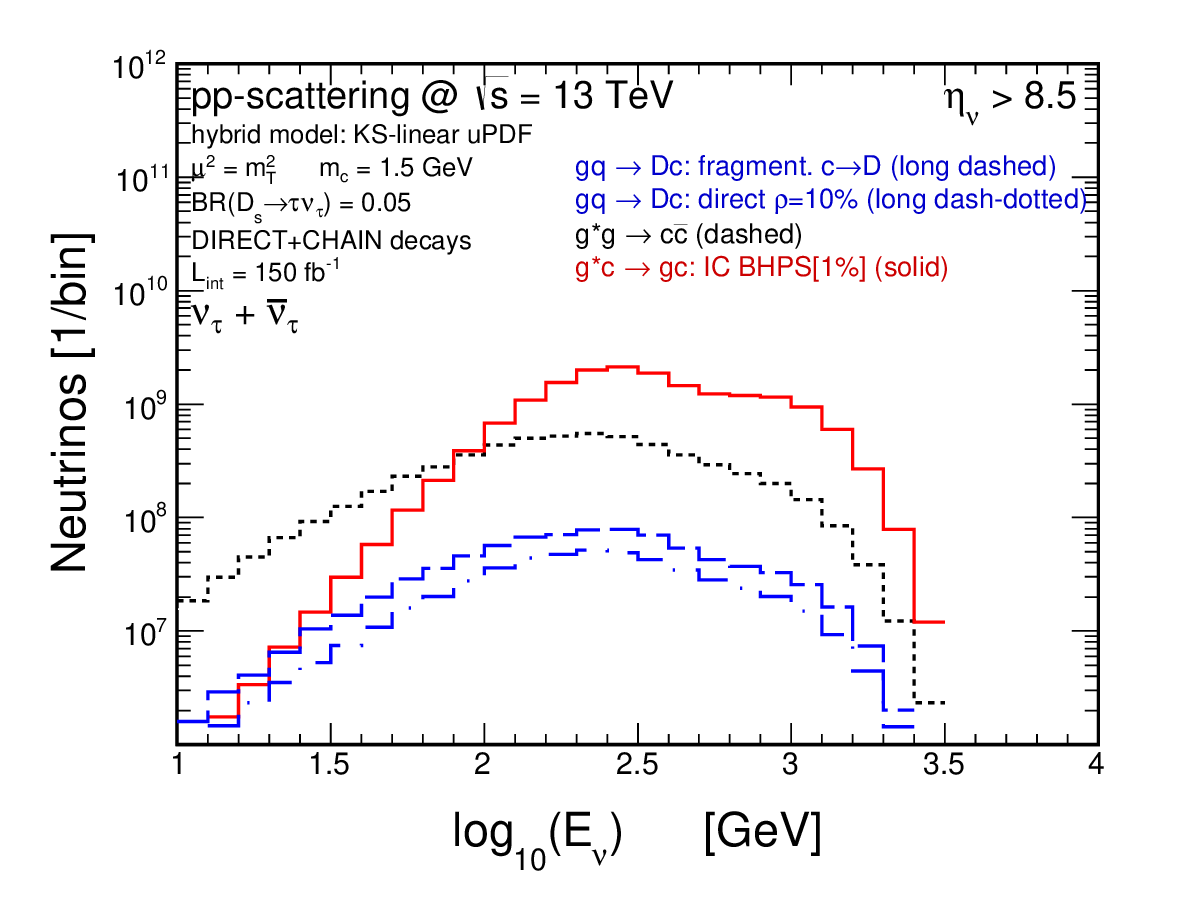}}
\end{minipage}
  \caption{
\small Energy distribution of tau neutrinos+anineutrinos for 
$\eta >$ 8.5 (FASER$\nu$).
}
\label{fig:tau_neutrinos}
\end{figure}

Another option to identify the subleading contributions 
is to investigate
energy distributions of $\nu_{\tau}$ neutrinos which are, however, difficult 
to measure experimentally.
Such distributions are shown in Fig.\ref{fig:tau_neutrinos}.
Here again the contribution of subleading mechanisms dominates over 
the traditional gluon-gluon fusion mechanism. 
In addition, there is no contribution of light mesons due to limited
phase space for $\tau$ production in the $D_s$ decay.
In this case the contribution due to recombination is small compared
to electron and muon neutrinos case because
$s(x) \ll u_{\mathrm{val}}(x), d_{\mathrm{val}}(x)$.
Therefore the measurement of $\nu_{\tau}$ and/or ${\bar \nu}_{\tau}$
seems optimal to pin down the IC contribution in the nucleon.

\section{Conclusions}

In this paper we have discussed forward production of charm quarks, 
charmed mesons and different types of neutrinos at the LHC energies.
Different mechanisms have been considered. The gluon-gluon fusion production of charm has been calculated within 
the $k_T$-factorization and hybrid approaches with different unintegrated gluon distributions
from the literature.

We have calculated transverse momentum distributions of 
$D^0 + {\bar D}^0$ for different bins of meson rapidities.
The results of the calculations have been compared to the LHCb
experimental data. A very good agreement has be achieved for
the Martin-Ryskin-Watt unintegrated gluon distribution for each bin of rapidity. The Kutak-Sapeta model
gives somewhat worse description of the LHCb data. Both models lead, however, to a similar predictions for far-forward charm production.

We have discussed the range of gluon longitudinal momentum
fractions for different ranges of rapidity. Already the LHC data
test longitudinal gluon momentum fractions as small as 10$^{-5}$.

We have shown that the intrinsic charm and recombination contributions
give rather negligible contribution for the LHCb rapidity range.
However, in very forward directions the mechanisms start to be crucial.
Using estimation of model parameters obtained recently
from the analysis of fixed-target data we have presented our
predictions for the LHC energy $\sqrt{s}$ = 13 TeV. 
Both the considered "subleading"
mechanisms win in forward directions with the dominant at midrapidity 
gluon-gluon fusion component. However, it is not possible to verify 
the prediction for the $D$ mesons experimentally.

We have calculated also different species of neutrinos/antineutrinos 
from semileptonic decays of different species of $D$ mesons. 
There is a good chance that 
energy distributions of $\nu_{e}$ neutrinos may 
provide some estimation on the subleading IC and recombination
mechanisms. However, the two subleading mechanisms compete.
The energy spectrum of $\nu_{\tau}$ neutrinos coming
from the decay of $D_s$ mesons may provide valueable information on 
the size of the intrinsic charm since here the recombination mechanism 
is reduced due to smallness of strange quark distributions.
In addition, in this case there is no conventional contributions related 
to semileptonic decays of pions and kaons.

In summary, we think that the measurement of energy distributions
of far-forward neutrinos/antineutrinos should provide new information
on subleading contributions to charm production, especially on the intrinsic
charm content of the proton.
Very large fluxes of high-energy neutrinos for FASER$\nu$ have been predicted.
A realistic simulation of the target for neutrino measurement should 
be performed in future.

\vskip+5mm
{\bf Acknowledgments}\

This study was supported by the Polish National Science Center grant 
UMO-2018/31/B/ST2/03537
and by the Center for Innovation and Transfer of Natural Sciences and Engineering Knowledge in Rzesz{\'o}w.


\end{document}